\documentclass[twocolumn]{aastex631}
\usepackage{natbib}
\usepackage{color}
\usepackage{hyperref} %----set up a hyperlink to citations, equations...
\def    \apjl  		{\rm {ApJL}}
\def    \apj  		{\rm {ApJ}}
\def    \mnras  	{\rm {MNRAS}}
\def    \araa  		{\rm {ARA\& A}}
\def    \apjl  		{\rm {ApJL}}

\def	\cm		{\,{\rm {cm}}}
\def	\K		{\,{\rm K}}
\def	\g		{\,{\rm {g}}}
\def	\mum	{\,{\mu \rm{m}}}

\def \bea {\begin{eqnarray}}
\def \ena {\end{eqnarray}}                  

\def\arcsec{\hbox{$^{\prime\prime}$}}

\def	\B	{{\rm B}}

\def	\C	{{\rm C}}

\def	\cm	{\,{\rm cm}}
\def	\km	{\,{\rm km}}
\def	\d	{{\rm d}}

\def	\D	{{\rm D}}

\def	\eV	{\,{\rm eV}\,}
\def	\g	{\,{\rm g}}

\def	\G	{{\rm G}}

\def	\H	{{\rm H}}

\def	\K	{{\rm K}}

\def	\N	{{\rm N}}

\def	\s	{\,{\rm s}}

\def	\V	{{\rm V}}

\def\kms{km\,s$^{-1}$}

\begin{document}

\title{Gas kinematics and dynamics of Carina Pillars: A case study of G287.76-0.87}

% Authors list
\author{Ngo Duy Tung}
\affiliation{University of Science and Technology of Hanoi, Vietnam Academy of Science and Technology, 18 Hoang Quoc Viet, Hanoi, Vietnam}
\affiliation{Université de Strasbourg, CNRS, Observatoire astronomique de Strasbourg, UMR 7550, 11 rue de l'Université, 67000, Strasbourg, France}
\affiliation{Université Paris Cité, Université Paris-Saclay, CEA, CNRS, AIM, F-91191, Gif-sur-Yvette, France}
\email{duy-tung.ngo@cea.fr}

\author{Le Ngoc Tram}
\affiliation{Max-Planck-Institut für Radioastronomie, Auf dem Hügel 69, 53-121, Bonn, Germany}

\author{Archana Soam}
\affiliation{Indian Institute of Astrophysics, II Block, Koramangala, Bengaluru 560034, India}
\affiliation{SOFIA Science Center, Universities Space Research Association, NASA Ames Research Center, Moffett Field, CA 94035, USA}

\author{William T. Reach}
\affiliation{SOFIA Science Center, Universities Space Research Association, NASA Ames Research Center, Moffett Field, CA 94035, USA}
\affiliation{Space Science Institute, 4765 Walnut St STE B, Boulder, CO 80301, USA}

\author{Edwin Das}
\affiliation{Department of Physics and Electronics, CHRIST (Deemed to be University), Bangalore, Karnataka 560034, India}

\author{Ed Chambers}
\affiliation{SOFIA Science Center, Universities Space Research Association, NASA Ames Research Center, Moffett Field, CA 94035, USA}
\affiliation{Space Science Institute, 4765 Walnut St STE B, Boulder, CO 80301, USA}

\author{Blesson Mathew}
\affiliation{Department of Physics and Electronics, CHRIST (Deemed to be University), Bangalore, Karnataka 560034, India}

\correspondingauthor{Le Ngoc Tram and Archana Soam}
\email{nle@mpifr-bonn.mpg.de and archana.soam@iiap.res.in}

\acceptjournal{ApJ}

\begin{abstract}
We study the kinematics of a pillar, namely G287.76-0.87, using three rotational lines of $^{12}$CO(5-4), $^{12}$CO(8-7), $^{12}$CO(11-10), and a fine structure line of [OI] $63\,\mu$m Southern Carina observed by SOFIA/GREAT. This pillar is irradiated by the associated massive star cluster Trumpler 16, which includes $\eta$~ Carina. Our analysis shows that the relative velocity of the pillar 
with respect to this ionization source is small, $\sim 1\,$\kms, and the gas motion in the tail is more turbulent than in the head. We also performed analytical calculations to estimate the gas column density in local thermal equilibrium (LTE) conditions, which yields $N_{\rm CO}$ as $(\sim 0.2 -5)\times 10^{17}\cm^{-2}$. We further constrain the gas's physical properties in non-LTE conditions using RADEX. The non-LTE estimations result in $n_{\rm H_{2}} \simeq 10^{5}\cm^{-3}$ and $N_{\rm CO} \simeq 10^{16}\,\rm cm^{-2}$. We found that the thermal pressure within the G287.76-0.87 pillar is sufficiently high to make it stable for the surrounding hot gas and radiation feedback if the winds are not active. While they are active, stellar winds from the clustered stars sculpt the surrounding molecular cloud into pillars within the giant bubble around $\eta$~Carina. 
 
\end{abstract}
\keywords{ISM: Clouds – infrared: ISM – ISM: lines and bands – ISM: individual (Carina) – ISM: molecules – (ISM:) photon-dominated region (PDR)}

\section{Introduction\label{sec:intro}}
\begin{figure}[!ht]
    \centering
    \includegraphics[width=0.5\textwidth]{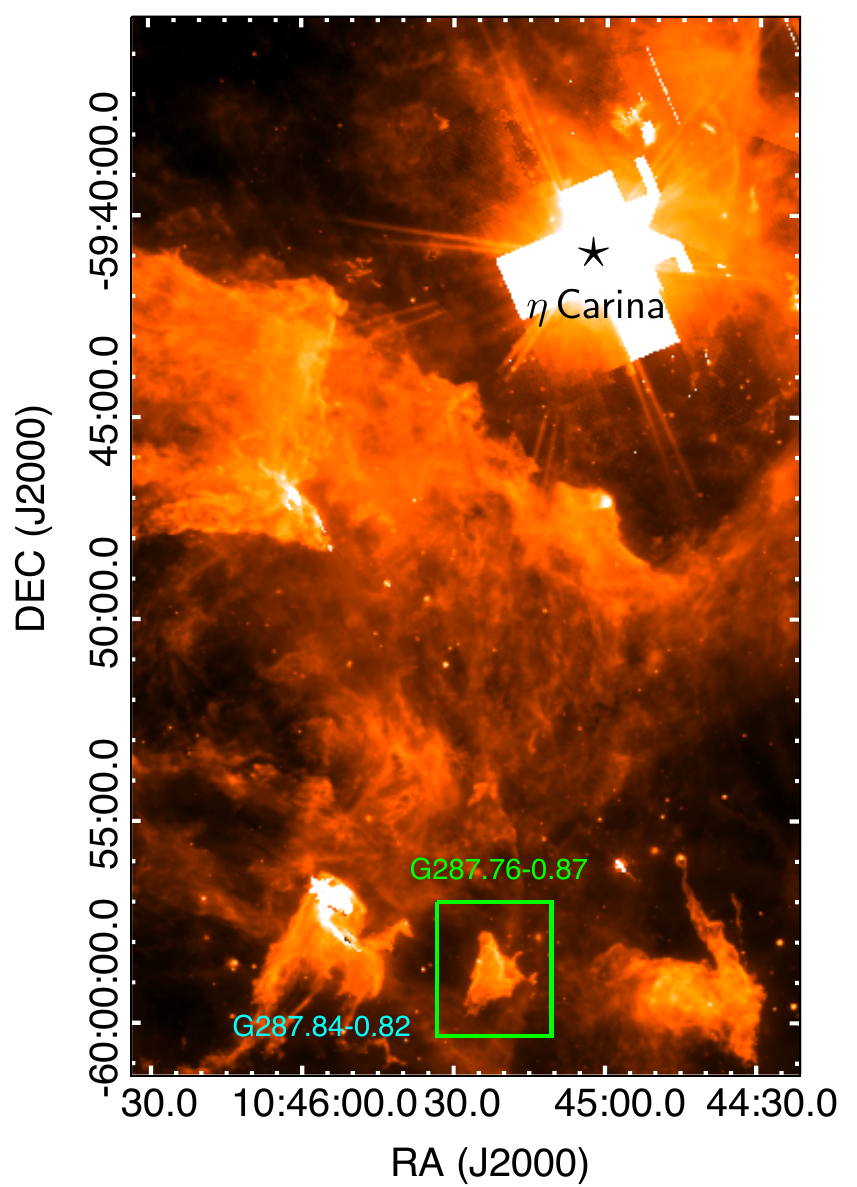}
    \caption{Southern Carina Nebula observed by {\it Spitzer} at $8\mum$. Our target (pillar G287.76-0.87) is highlighted by a green box and located next to the Treasure Chest pillar (G287.84-0.82, \citealt{2005AJ....129..888S}). The ionization source is the {\bf Tr16 cluster}, including the supermassive star $\eta$ Carina, shown by a star symbol. The area around $\eta$ Carina is white because it saturates the $8\,\mu$m camera.}
    \label{fig:Spitzer}
\end{figure}

The Carina Nebula, one of the largest and brightest diffuse nebulae in the southern sky, is home to many massive star-forming processes. In this region of star formation, the most massive stars (O/B stars) generate intense ultraviolet (UV) radiation of energy $E>13.6\eV$ as feedback to the nascent molecular cloud (see Figure \ref{fig:Spitzer}). These energetic UV photons ionize atomic hydrogen (H) and create an "\textsc{Hii} region" (see, e.g., \citealt{1939ApJ....89..526S}; \citealt{1954BAN....12..187K}) around the stars. At the boundaries of the \textsc{Hii} region, the UV photons that are less energetic than the ionization potential of H (that is, $E < 13.6\eV$) escape and penetrate the surrounding molecular gas. Along the way, they dissociate molecules, heat dust grains, and ionize atoms such as carbon, silicon, and sulfur, which have ionization potentials lower than $13.6\eV$. This gives rise to a region where H is neutral, but low ionization potential atoms are ionized, and molecules such as CO are photodissociated, which is called a photodissociation region or PDR. (\citealt{1997ARA&A..35..179H}). 

\begin{figure}[!ht]
    \centering
    \includegraphics[width=0.5\textwidth]{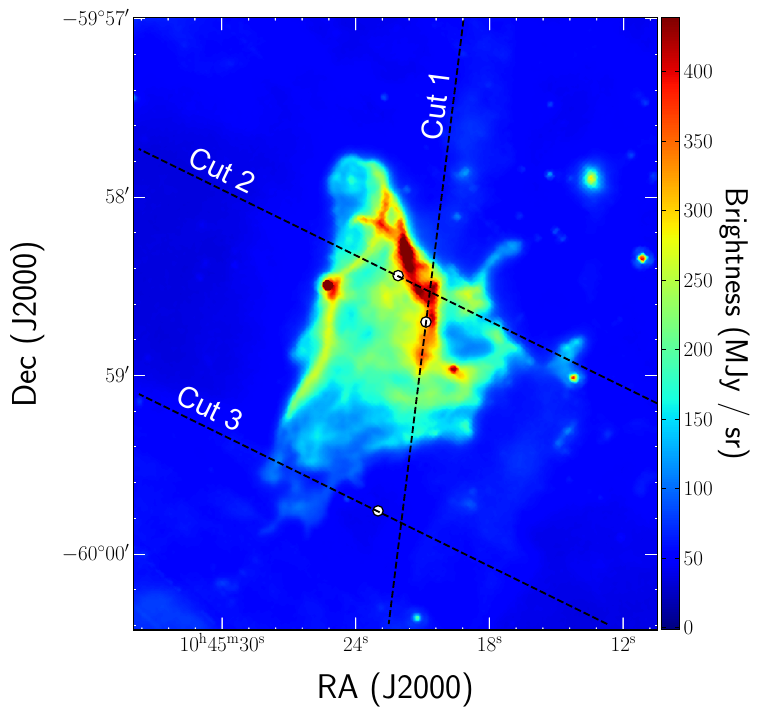}
    \caption{Zoom on the pillar G287.76-0.87: The dashed lines indicate the cuts for position-velocity diagrams presented in Section \ref{subsec:PV} and the white circles mark their corresponding zero positions.}
    \label{fig:cut_PV}
\end{figure}

At the interface between a molecular cloud and an \textsc{Hii} region, there exist some of the most spectacular features, namely {\it globules} and {\it pillars}. Pillars, which arise when UV radiation from hot stars propagates into a dense, clumpy region in the surrounding molecular cloud (\citealt{1996AJ....111.2349H}), have an elongated column-like shape and a physical connection to the gas reservoir of the cloud. Globules, which result from the separation of pillars from the gas (\citealt{1986A&A...156..301W}; \citealt{1993A&A...274L..33W}; \citealt{2004A&A...414.1017T}; \citealt{2012A&A...542L..18S}), are isolated (often with a young star located near the head) and have a head-tail structure pointing toward the illuminating source. In pillars and globules, the gas is strongly influenced by radiation; thus, the absorption of UV photons of $E < 13.6\eV$ by the gas leads to intense emission of [\textsc{Cii}] at $158\mum$, [\textsc{Oi}] at $63, 146 \mum$, and H$_2$ ro-vibrational transitions. Deeper in PDRs, CO rotational and [\textsc{Ci}] $370, 609 \mum$ lines can be observed. Therefore, observations of these spectral lines in high spectral resolution (\citealt{2019A&A...626A.131M}; \citealt{2019ApJ...882...11A}; \citealt{2020MNRAS.497.2651K}) can help us to trace the gas kinematics and understand the physical conditions of the gas in the emitting region.

The energy injection from an intense ionization source plays both positive and negative roles in the surrounding clouds. This feedback could trigger a new generation of stars to form (\citealt{1977ApJ...214..725E}) or halt this process or even disrupt the cloud structure (\citealt{pabst19}). In the case of Southern Carina, \cite{2004A&A...418..563R} showed that a cluster of young stellar objects (YSOs) are likely to form in the pillar G287.84-0.82 (as known as Treasure Chest, see Figure \ref{fig:Spitzer}), but there is no evidence of star formation in the neighboring pillar (G287.76-0.87, highlighted by a box in Figure \ref{fig:Spitzer}). 

In this work, we will focus on the latter pillar, which is shown in Figure \ref{fig:cut_PV}. We aim at characterizing the kinematic structure of this cloud using the very high spectral resolution of mid-Infrared rotational transitions of CO and the fine-structure line of oxygen observed by the GREAT instrument (\citealt{2018JAI.....740014R}) aboard the Stratospheric Observatory for Infrared Astronomy (SOFIA, \citealt{young12}). Furthermore, being the cooling lines of PDRs, these lines are ideal tracers to constrain the pillar's physical properties.

This paper is structured as follows. Our study's observations and data acquisition are presented in Section \ref{sec:obs}. The results obtained and the analyses applied to the results are presented in Section \ref{sec:result}. We discuss our results in Section \ref{sec:discussion}. Finally, a summary and conclusion of our findings are given in Section \ref{sec:sum}.

\section{Observations and data acquisition} \label{sec:obs}

The observations used in this work are made using The German REceiver for Astronomy at Terahertz Frequencies (GREAT) onboard
SOFIA, in its 4GREAT/HFA (High Frequency Array) configuration. 
GREAT is able to observe five frequencies simultaneously. The 4GREAT channel, which consists of four co-aligned pixels at four frequencies (\citealt{duran21}), was tuned to $^{12}$CO(5-4), $^{12}$CO(8-7), $^{12}$CO(11-10), and $^{12}$CO(22-21). The HFA, which consists of seven independent pixels in a hexagonal geometry with a central beam, was tuned to the [\textsc{Oi}] line at $63\mum$ at 4.7 THz.

Observations of G287.76-0.87 were made during the 2019 SOFIA deployment to New Zealand as part of the Director's Discretionary Time (DDT) project (75\_0038). The data were obtained in about 130 minutes, spread across two flights on June 26 and 27, 2019. The maps were obtained in single beam-switch mode and designed to fully sample the 6.3\arcsec [\textsc{Oi}] beam, with spectral readouts every 3\arcsec. Each readout had 0.4 seconds of on-source integration time. For the [\textsc{Oi}] map, each of the 7 pixels creates its own independent map, resulting in about 2.8 seconds of on-source time per map point. For the 4GREAT maps, the 3\arcsec\, spacing resulted in over-sampled maps. During the observations, the 4G4 channel that was tuned to CO(22-21) was unstable, and therefore those data are not included in our analysis. 

The data were reduced and calibrated by the GREAT team using their {\it kalibrate} software package (\citealt{2012A&A...542L...4G}) for atmospheric calibration and GILDAS/CLASS\footnote{http://www.iram.fr/IRAMFR/GILDAS} for final beam calibration, baseline removal, and map convolution and gridding. The final maps have pixel sizes of 18\arcsec, 8.6\arcsec, 6.6\arcsec, and 2\arcsec\, for the CO(5-4), CO(8-7), CO(11-10), and [\textsc{O i}] maps, respectively. The spectra were all smoothed from their native resolution to channel widths of $\sim 0.3 \km \s^{-1}$. All data are publicly available at the Infrared Science Archive (IRSA\footnote{https://www.sofia.usra.edu/node/2684}).

\begin{figure*}
\centering
\includegraphics[width=\textwidth]{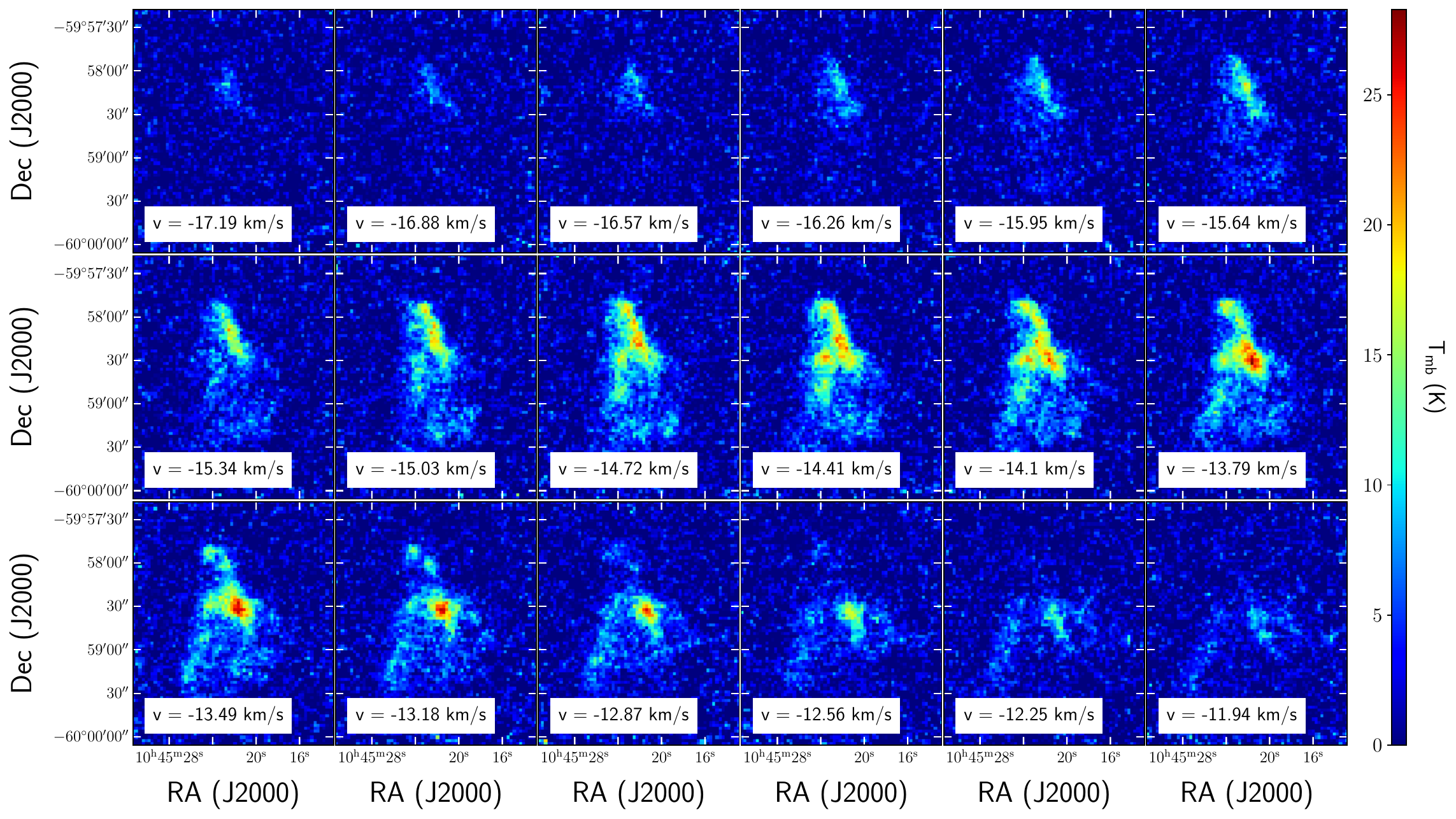}
\caption{Channel maps of [\textsc{O i}] $63\mum$, the highest spatial resolution in our data set. The motion is blue-shifted, indicating that the pillar is in front of the radiation source ($\eta$~Carina).}
\label{fig:channelmap}
\end{figure*}

\section{Results} \label{sec:result}
\subsection{Channel Maps}
We investigated the motion of gas in the pillars using velocity channel maps. Figure \ref{fig:channelmap} shows the channel maps of [\textsc{Oi}] 63$\,\mum$, whose spatial resolution is the highest in our data set. The channel of the CO lines will be shown later in Figures \ref{fig:channelmap_app2}-\ref{fig:channelmap_app3} in Appendixes. As the velocity increases, the signal becomes stronger, peaks at -14.7$\,\rm km\,s^{-1}$, and then starts to fade. The velocity for all four lines is negative, which means that the gas motion is blue-shifted with respect to the local standard of rest (LSR) by  $\sim -14 \km \s^{-1}$. This velocity is red-shifted with respect to the nearby cloud in $\eta$ car that has the LSR velocity $\sim -20\,$\kms. We discuss the motion of the pillar in Section \ref{sec:pillar_motion}. The southwest region of the gas seems very peculiar in [OI] 63$\mum$, which is not seen in CO transitions (Figures \ref{fig:channelmap_app2},\ref{fig:channelmap_app3}) owing to the larger beam sizes of the CO observations. 

\subsection{Moment maps}
\begin{figure*}
\includegraphics[width=0.5\textwidth]{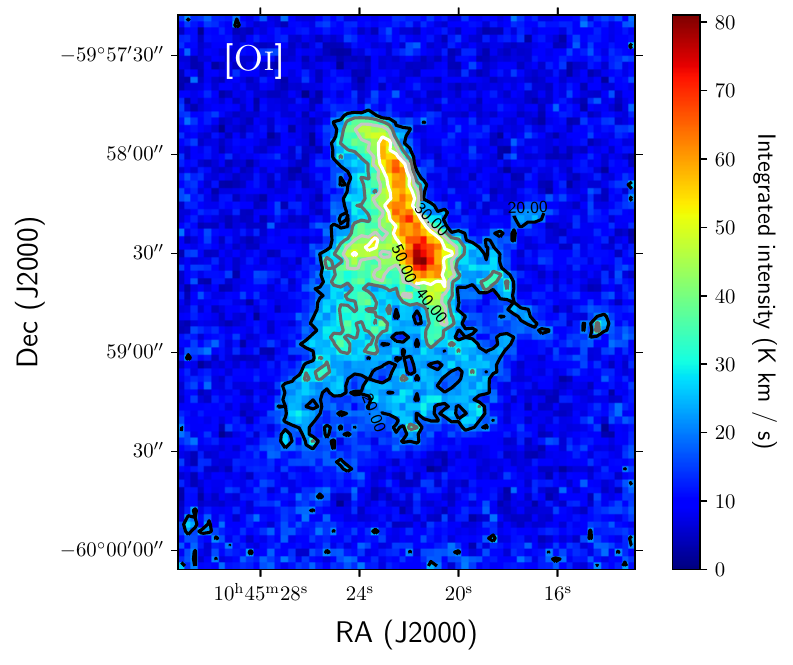}
\includegraphics[width=0.5\textwidth]{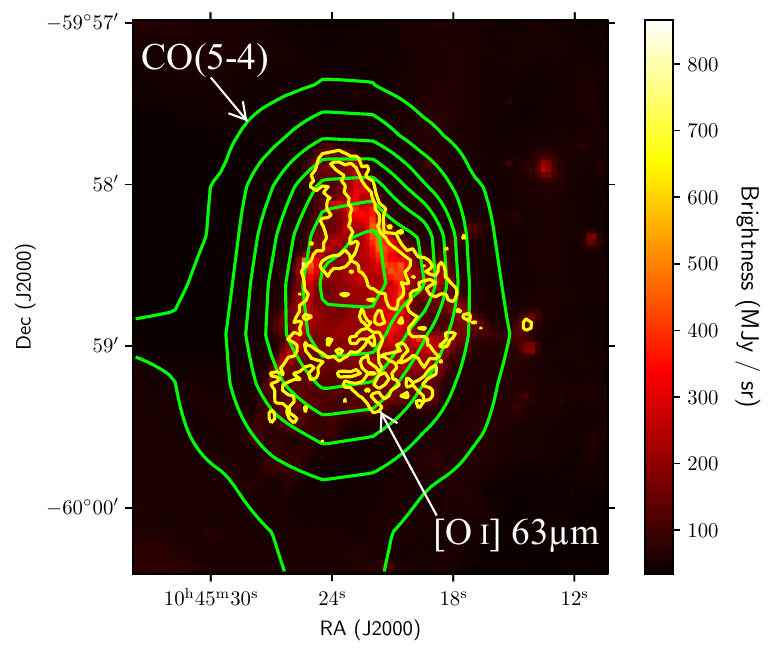}
\includegraphics[width=0.5\textwidth]{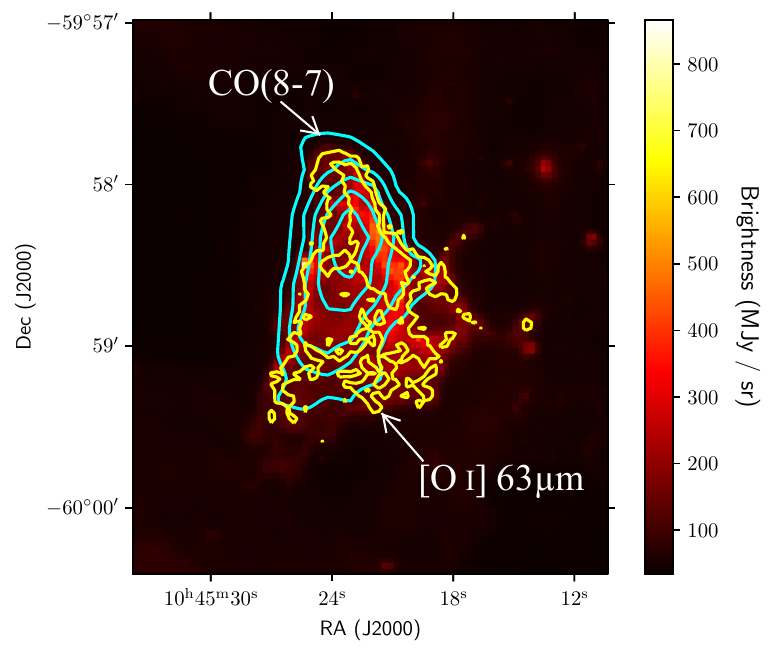}
\includegraphics[width=0.5\textwidth]{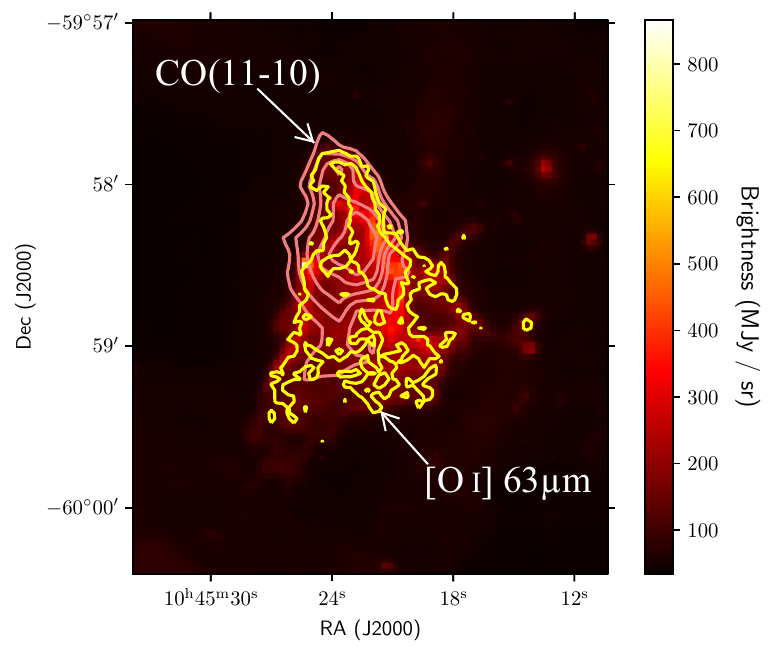}
\caption{Integrated intensity maps of [\textsc{Oi}] is shown in the upper left panel. Other panels show the correlation between the [\textsc{Oi}], CO(5-4), CO(8-7), CO(11-10) integrated intensities (contours) and the $Spitzer$ 8$\mu$m map (background color).}
\label{fig:moment0}
\end{figure*}

The upper left panel of Figure \ref{fig:moment0} shows the integrated brightness map of the [\textsc{Oi}] $63\mum$ line. The map shows that the emission peaks at the northwest of the pillar and decreases radially along the northwest-to-southeast direction. This can be reconciled with the fact that the northwest region of the pillar is exposed to the radiation source ($\eta$~Carina). Radiation and interaction with the expanding \textsc{Hii} shell can excite the molecular and atomic gas, and excitation becomes radially weaker behind this interaction front.

Radiation and the expanding shell \textsc{Hii} also heat dust grains. Therefore, we compare the spatial correlation between gas emission and dust continuum in the pillars G287.76-0.87. The results are shown in Figure \ref{fig:moment0} (upper right and lower panels). The contours are the integrated intensity of [\textsc{Oi}] (yellow) and CO (other colors) emissions. The background represents the $8\mum$ {\it Spitzer} map,  which can trace the emission from polycyclic aromatic hydrocarbon (PAHs) molecules. The [\textsc{Oi}] $63\mum$ emission correlates very well with the 8 $\mu$m PAHs continuum emission and is located ahead of the CO lines, representing the PDR structure.

\begin{figure*}
\includegraphics[width=0.5\textwidth]{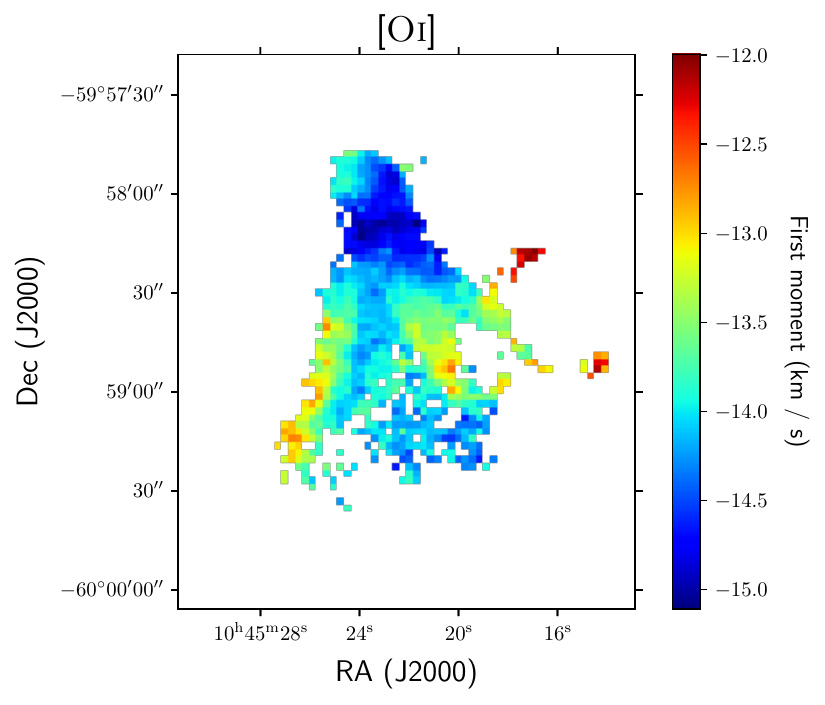}
\includegraphics[width=0.5\textwidth]{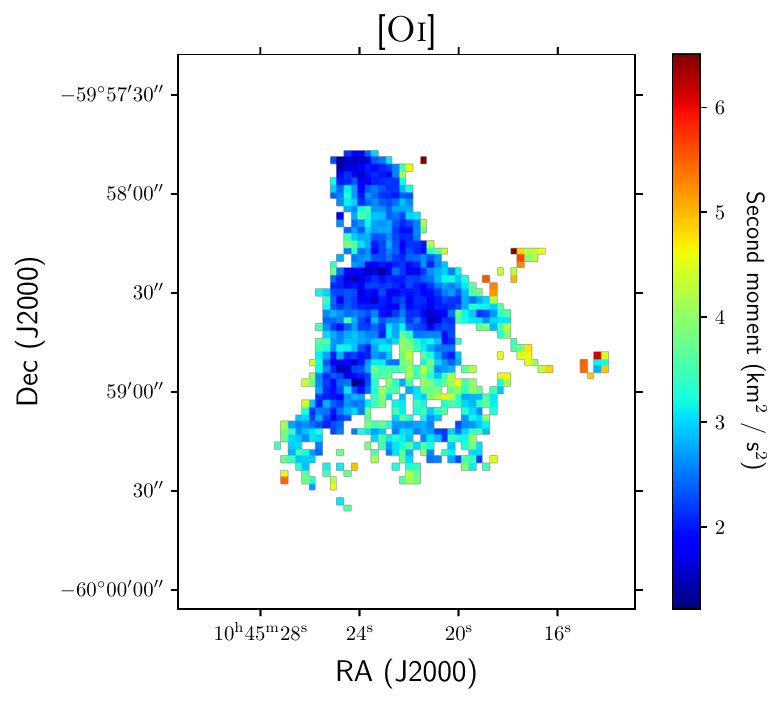}
\caption{
\textit{\textbf{Left}}: First-moment map of
[\textsc{Oi}] $63\mum$ integrated in velocity according to the channel as in Figure \ref{fig:channelmap}. Along the LOS, the head velocity is more negative than the tail velocity. \textit{\textbf{Right}}: Corresponding second-moment map of [\textsc{O i}]. The motion projected on the LOS in the tail is shown to be more turbulent than in the head.}
\label{fig:moment1}
\end{figure*}

The left panel of Figure \ref{fig:moment1} shows the first moment map of the [\textsc{Oi}] $63\mum$ lines, in units of \kms. The maps for the CO lines are shown in the Appendix (Figure \ref{fig:moment01_app}). To reduce noise, a mask is applied to these maps to select the data within the zeroth moment contours as shown in Figure \ref{fig:moment0}. Along the line of sight, the gas velocity at the head is more negative than that at the tail of the pillar.

In the right panel of Figure \ref{fig:moment1}, we show the second moment map of [\textsc{Oi}] 63 $\mu$m, in units of $\km^{2}\,\s^{-2}$. Here we choose not to show the moment maps for the other CO tracers because of their low spatial resolutions, which result in little meaningful information. Interestingly, although having a lower intensity-weighted velocity as suggested by the first-moment maps, the tail's motion is more turbulent than that of the head, with a maximum dispersion of around $1-3\,$\kms. A very similar effect effect has recently been noted in stellar wind shells, specifically the Orion Veil \citep{kavak22} and RCW 120 \citep{kirsanova23}.

\subsection{Position-velocity diagrams}\label{subsec:PV}

With all information from the channel maps and moments maps, we can reconstruct a 3D motion of the pillar if we know how the velocity is related to the position. In this section, we aim at that goal by means of position-velocity (PV) diagrams with a few slices through the pillar. Using the Python module \texttt{pvextractor} (\citealt{2016ascl.soft08010G}), we extract the PV diagrams in CO(5-4), CO(8-7), CO(11-10), and \textsc{[O i]} $63\mum$ along the three cuts shown in the right panel of Figure \ref{fig:cut_PV}. 

The first cut is set to point toward the $\eta$~Carina. The zero position for this cut is at ($10^{\rm h} 45^{\rm m} 21^{\rm s}$, $-59^{\circ}58.7'$). Apparently, CO(5-4), CO(8-7), and [\textsc{Oi}] 63$\,\mum$ show a clear structure, whereas CO(11-10) does not because this transition does not trace well in this direction (see Figure \ref{fig:moment0}, bottom right panel). Signals rise sharply in the northwest and narrow down in the southeast.   

The second cut is perpendicular to the first. The zero position for this cut is at ($10^{\rm h} 45^{\rm m} 22^{\rm s}$, $-59^{\circ}58.44'$). CO(5-4) seems to be compact with little velocity spread, and the CO(8-7) shows that the gas motion is not highly dispersed, yet the spatial spread of the gas at the same velocity is quite visible. On the other hand, CO(11-10) lines are broader at this position, with FWHM of 4$\,$\kms. The gas motion spans a range of $\sim 5\,$\kms, which is larger than those of other tracers. The plot for [\textsc{Oi}] 63$\,\mum$ in this cut does not provide much information. It seems to have much noise compared to the main signal.

The third cut, whose zero position is at ($10^{\rm h} 45^{\rm m} 23^{\rm s}$, $-59^{\circ}59.75'$), is made based on the observation that there is a strange feature (a hole) at the southeast of the clump. In this cut, the velocity component is not seen by CO(11-10) and [\textsc{Oi}] 63 $\mum$, suggesting that the gas is more diffuse than the rest of the clump, as we saw in the \textit{Spitzer} image (Figure \ref{fig:cut_PV}).  

\begin{figure*}
\centering
\includegraphics[width=\textwidth]{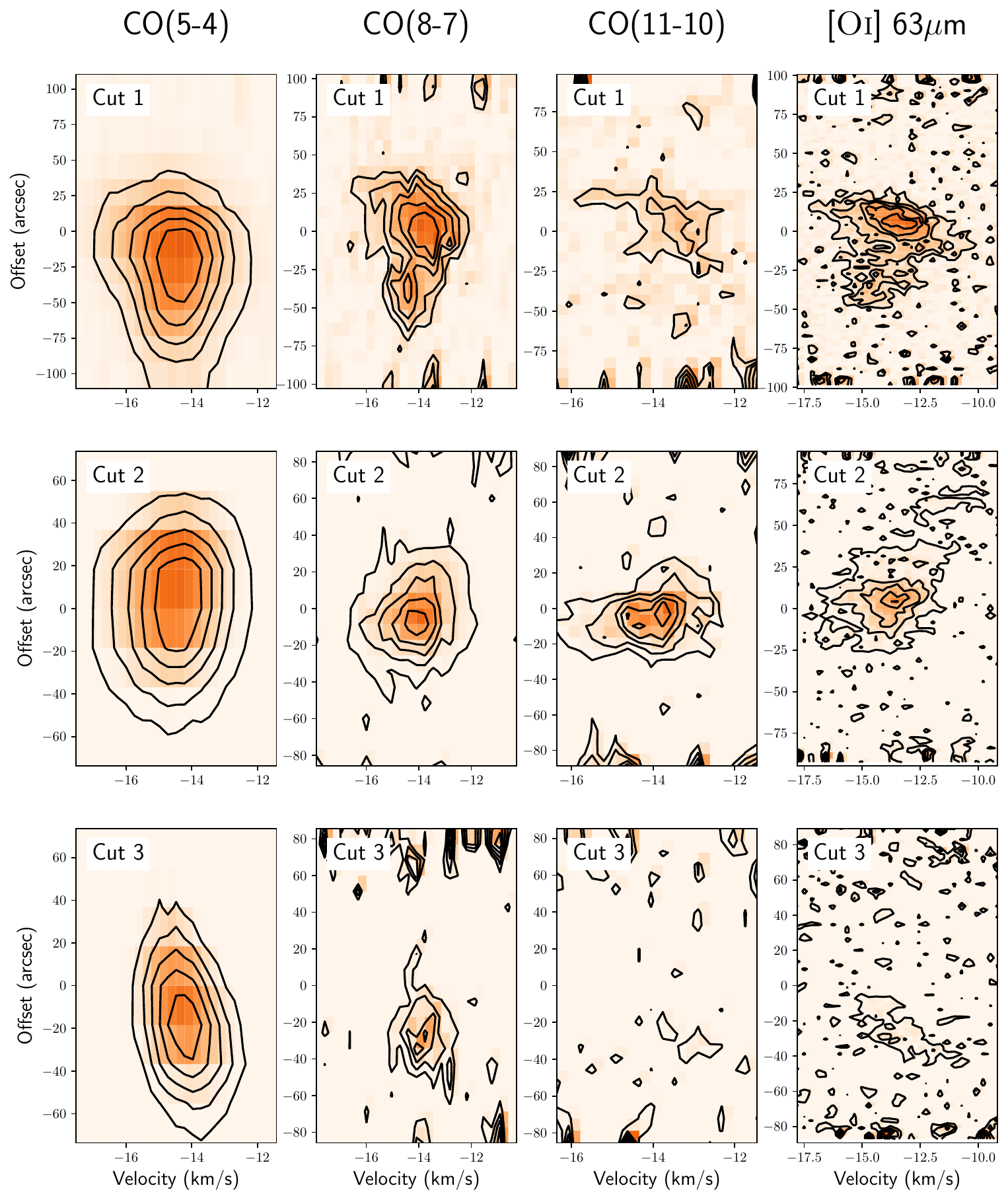}

\caption{Position–velocity plots of CO(5-4), CO(8-7), CO(11-10), and [\textsc{Oi}] in contours enhanced with orange scale along the cuts shown in Figure \ref{fig:cut_PV}. Five contours from 3K to the peak value of the brightness temperature are plotted for each cut. The orange scale spans from zero to 1.5 times the maximum temperature.}
\label{fig:PV}
\end{figure*}

\subsection{Spectral distribution}
\begin{figure*}
\includegraphics[width=0.5\textwidth]{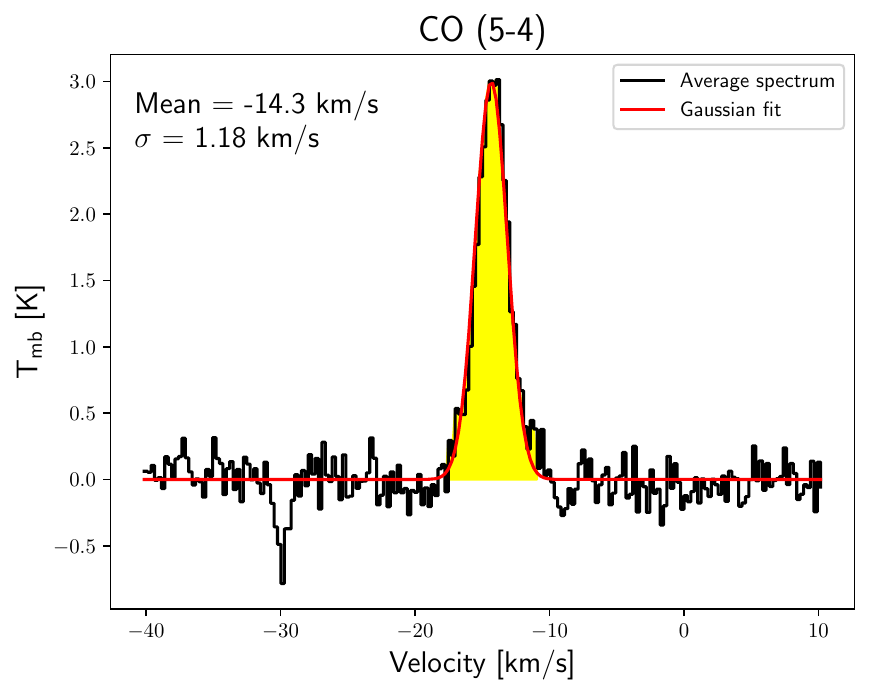}
\includegraphics[width=0.5\textwidth]{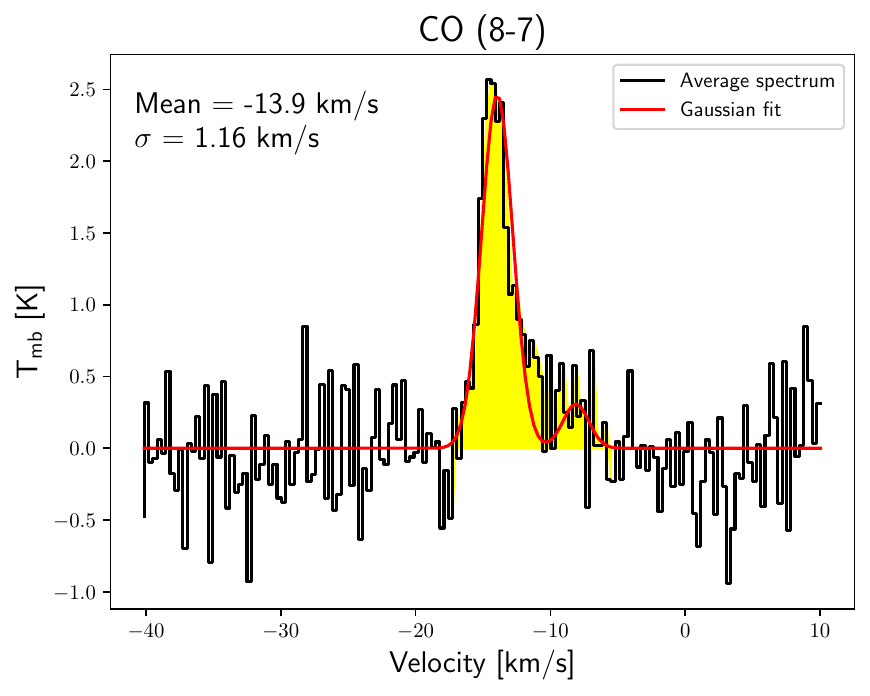} \\
\includegraphics[width=0.5\textwidth]{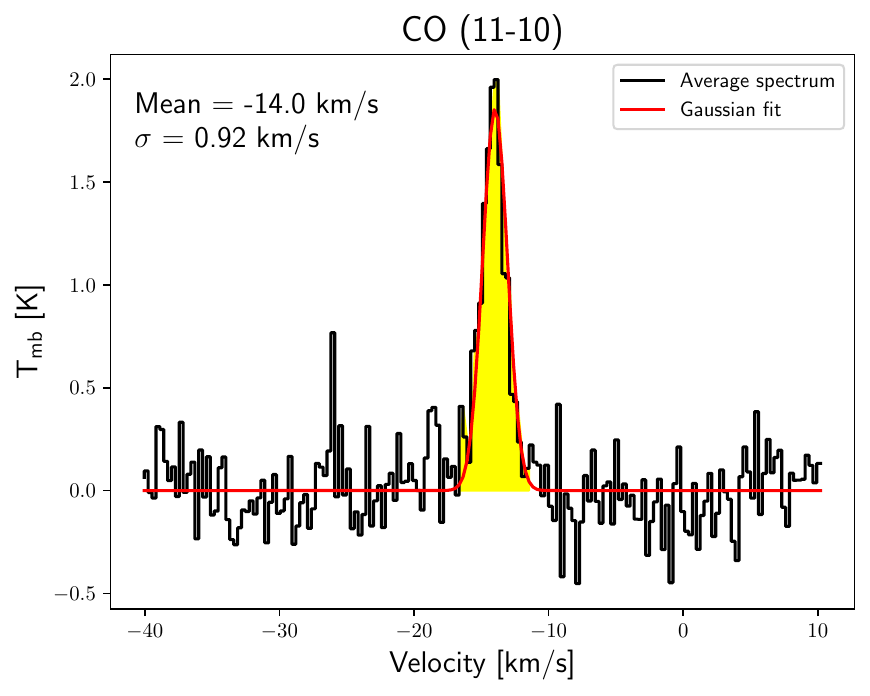}
\includegraphics[width=0.5\textwidth]{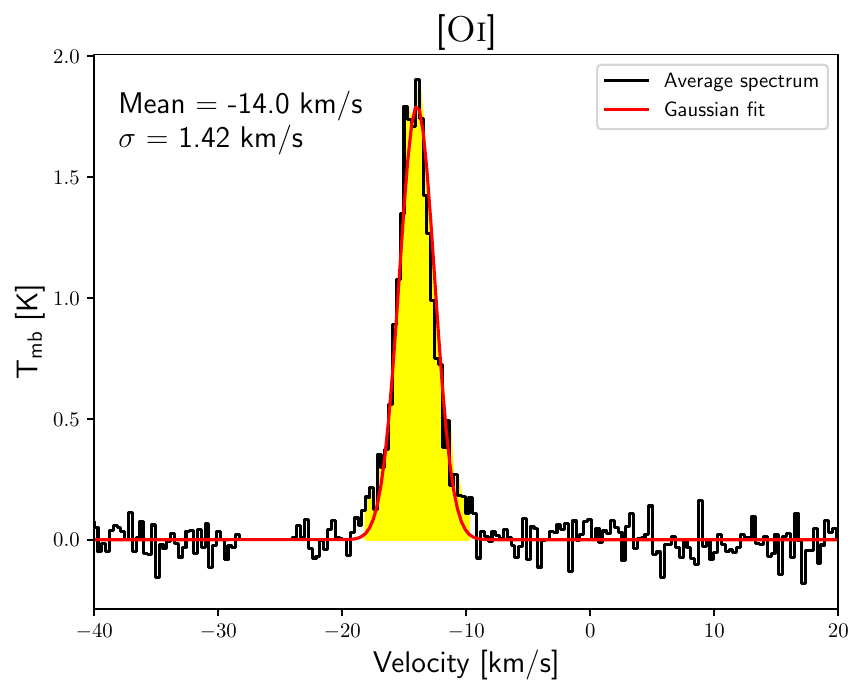}
\caption{Average spectra of four transition lines with the fitted Gaussians (red curves). The mean and standard deviation (of the main Gaussian) are noted in each panel. The yellow-filled area corresponds to the signal range, which lies within $[\mu-3\sigma,\mu+3\sigma]$.}
\label{fig:spectra}
\end{figure*}

Figure \ref{fig:spectra} shows the average spectra for all the lines. 
The velocity resolution  of 4GREAT is finer than 0.1$\,$\kms \citep{duran21}, which well resolves the spectral lines.
We then fit these spectra with a single Gaussian function. The corresponding mean $\mu$ and standard deviation $\sigma$ are marked in each panel. For each line, our confident signal is selected between $\mu-3\sigma$ and $\mu+3\sigma$, which accounts for 99.7\% of the signal. For the CO(8-7) line, since a single Gaussian function does not seem to cover all the signal, we try to fit it with a double Gaussian function. The parameters for the second Gaussian are $\mu = -8.1\km\,\s^{-1}$ and $\sigma = -0.97\km\,\s^{-1}$. The nature of this second component needs to be determined in future studies. From these fitted parameters, one can see that the [\textsc{Oi}] line is broader than the CO lines.

\subsection{Physical properties}

Our goal in this section is to constrain some key physical properties of the pillar G287.76-0.87, i.e., the kinetic temperature $T_{\rm kin}$, the number density $n_{\rm H_{2}}$ and the column density $N_{\rm CO}$ (which can convert to the gas column density), using the observed spectra. However, since CO transitions are either optically thick or optically thin, and we do not have data for other isotopes of CO (e.g., $^{13}$CO, which is optically thin) and/or C$^{+}$ (e.g., $^{12}$C$^+$, $^{13}$C$^{+}$), which are typical tracers for PDR study, toward our target. Therefore, we first propose a different analytical approach to estimate the gas column density in local thermal equilibrium (LTE) conditions and then perform the non-LTE modeling using the radiative transfer program RADEX.

\subsubsection{LTE calculations}
Our dataset contains only three transitions of the $^{12}$CO and one atomic [\textsc{Oi}] transition. The $^{12}$CO transitions can be either optically thick or optically thin, whereas the [\textsc{Oi}] 63$\,\mu$m seems to be thick in the PDR (see \citealt{2019ApJ...887...54G}). In this section, we, therefore, estimate the physical properties from the line ratios of $^{12}$CO in LTE conditions by assuming that they are either optically thin or optically thick, as introduced in \textcite{2012ApJS..203...23H}, conditioning all the lines are tracing the medium with the same excitation temperature, and their filling factors are all equal. 

Let us consider a rotational transition at frequency $\nu$. Its brightness temperature in LTE (i.e., the source function is approximated as a black-body function) is
\bea
    T_{\rm B} = f [J_{\nu}(T_{\rm ex}) - J_{\nu}(T_{\rm bg})][1-\exp(-\tau_{\nu})],
\ena
where $f$ is the filling factor, $T_{\rm ex}$ is the excitation temperature, $T_{\rm bg}=2.73\,\K$ is the background temperature, $\tau$ is the optical depth, and $J_{\nu}$ is the black-body temperature defined at temperature $T$ as
\bea
    J_{\nu}(T) = \frac{h\nu/k_{\B}}{\exp(h\nu/k_{\B}T)-1},
\ena 

In the first case where both lines are optically thick ($\tau_{\nu}\gg 1$), the ratio of the integrated intensity of line 1 ($\nu_1$, $\tau_{\nu_{1}}$) over that of line 2 ($\nu_2$, $\tau_{\nu_{2}}$) is
\bea
    R_{\nu_1/\nu_2} = \frac{[J_{\nu_1}(T_{\rm ex})-J_{\nu_1}(T_{\rm bg})]}{[J_{\nu_2}(T_{\rm ex})-J_{\nu_2}(T_{\rm bg})]}.
\ena

Second case, both lines are optically thin ($\tau_{\nu} \ll 1$), which yields
\bea
    R_{\nu_1/\nu_2} = \frac{[J_{\nu_1}(T_{\rm ex})-J_{\nu_1}(T_{\rm bg})]\tau_1}{[J_{\nu_2}(T_{\rm ex})-J_{\nu_2}(T_{\rm bg})]\tau_2}
\ena
with $\tau$ being the opacity, and the opacity ratio
\bea
    \frac{\tau_1}{\tau_2}= \frac{\nu^{2}_{2}}{\nu^{2}_{1}}\frac{g_{u1}}{g_{u2}}\frac{A_{1}}{A_{2}}\frac{N^{\rm tot}_{1}}{N^{\rm tot}_{2}}\frac{Q_{2}}{Q_{1}}\frac{e^{-E_{l1}/T_{\rm ex}}}{e^{-E_{l2}/T_{\rm ex}}}\left[\frac{1-e^{-h\nu_1/k_{\B}T_{\rm ex}}}{1-e^{-h\nu_2/k_{\B}T_{\rm ex}}}\right],
\ena
where $A_{1}$ and $A_{2}$ are the Einstein spontaneous emission coefficients, $g_{u1}$ and $g_{u2}$ are the statistical weights of the upper levels, $E_{l1}\,(\K)$ and $\ E_{l2}\,(\K)$ are the excitation energies of the lower levels, $Q_{1}$ ($N^{\rm tot}_{1}$) and $Q_{2}$ ($N^{\rm tot}_{2}$) are the partition functions (total column density) of the molecule/atom for line 1 and line 2, respectively. Therefore, if the two lines are from the same molecule/atom transitions, the opacity ratio is independent of the partition function and the total column density.     

The third case, the transition $\nu_1$ is optically thin, while the transition $\nu_2$ is optically thick. The integrated intensity ratio is
\bea
    R_{\nu_1/\nu_2} = \frac{[J_{\nu_1}(T_{\rm ex})-J_{\nu_1}(T_{\rm bg})]}{[J_{\nu_2}(T_{\rm ex})-J_{\nu_2}(T_{\rm bg})]}\tau_1.
\ena

\begin{figure*}
    \includegraphics[width=\textwidth]{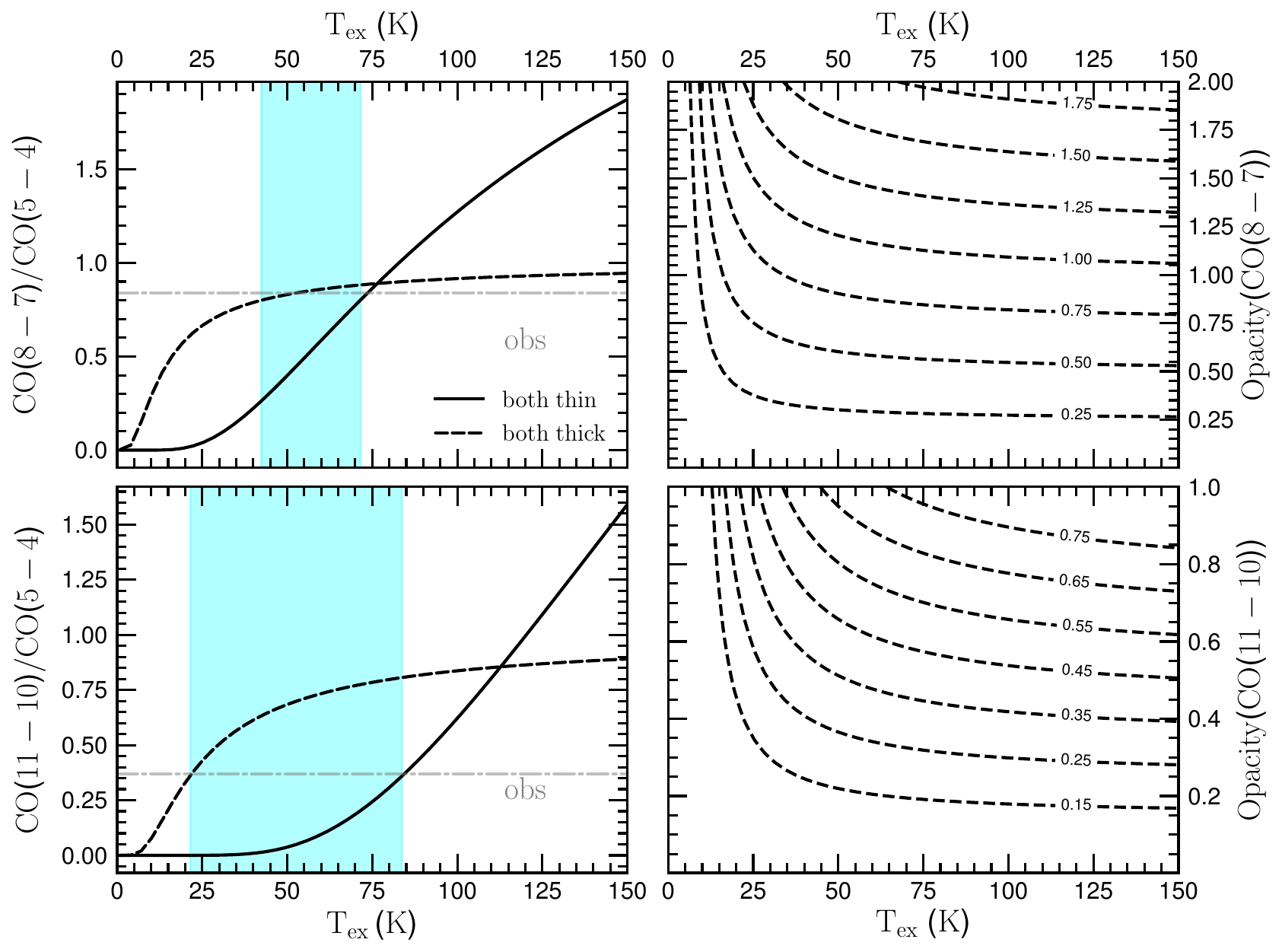}
    \caption{\textbf{\textit{Left}}: The ratio of the integrated intensity of CO(8-7)/CO(5-4) (top) and of CO(11-10)/CO(5-4) (bottom) as a function of the excitation temperature when they are both optically thin (solid line) and both optically thick (dashed line). \textbf{\textit{Right}}: Their integrated intensity ratios are computed as a function of the excitation temperature and the opacity, assuming that CO(5-4) is optically thick. Dash-dotted lines mark the observed ratio calculated from the average integrated intensity of the lines.}
    \label{fig:LTE_conditions}
\end{figure*}

Figure \ref{fig:LTE_conditions}, left panels show the results for the ratios CO(8-7)/CO(5-4) and CO(11-10)/CO(5-4) as a function of $T_{\rm ex}$, assuming these two lines are both optically thin (solid line) and optically thick (dashed line). The latter case slowly changes or is likely constant for $T_{\rm ex}\geq 40\K$. The horizontal lines show the observed ratios, which are CO(8-7)/CO(5-4) $\sim 0.8$ and CO(11-10)/CO(5-4) $\sim 0.35$. The range of $T_{\rm ex}$ can be determined by the intersection of this line and these two extreme cases above (cyan-filled region), which is not far from the dust temperature in this region ($\simeq 30\K$, see \citealt{2013A&A...554A...6R}). If we assume that these three CO lines trace the same excitation temperature, the $T_{\rm ex}$ spans from $40-70\K$. Figure \ref{fig:LTE_conditions}, the right panel shows the line ratio (contour lines) as a function of both $T_{\rm ex}$ and $\tau$ by assuming CO(5-4) is optically thick and CO(11-10) and CO(8-7) are optically thin. For $T_{\rm ex} \sim 40 - 70\,\K$, $\tau_{\rm CO(8-7)} \sim 0.8$, and $\tau_{\rm CO(11-10)} \sim 0.4$. We use $\tau_{\rm CO(11-10)} \sim 0.4$ because it likely manifests our assumption to compute the total density of CO gas which results in $N_{\rm CO}\sim 2\times10^{16}-5\times10^{17}\cm^{-2}$. We rule out the case of CO(5-4) optically thin, while CO(8-7) and CO(11-10) are optically thick (not shown here) because the optical depth returns to be greater than unity, which contradicts our assumption.

\subsubsection{Non-LTE with RADEX}
For the case of non-LTE conditions, we model the integrated intensity for each transition line of CO by the radiative transfer program RADEX (\citealt{2007A&A...468..627V}). To constrain $T_{\rm kin}$, $n_{\rm H_{2}}$, and $N_{\rm CO}$, we vary these three parameters at the same time and fit the diagram of the integrated intensity (I) vs. the quantum upper level (J). 
We use a Bayesian approach to perform the fits. The parameter space is investigated using an ensemble of Markov chains following the affine-invariant Markov chain Monte Carlo (MCMC; \citealt{2010CAMCS...5...65G}) algorithm. In particular, inspired by \cite{2017A&A...608A.144Y}, we couple the \texttt{emcee} (\citealt{2013PASP..125..306F}) library with a Python wrapper of RADEX called \texttt{pyradex}\footnote{https://github.com/keflavich/pyradex} to obtain RADEX results for each iteration and sample the posterior probability distribution function (PDF) with \texttt{emcee}. We use a total of 400 walkers, iterated with 2000 ``burn-in" steps followed by 10000 main steps, to sample the posterior PDF.

\begin{figure}
\centering
\includegraphics[width=0.5\textwidth]{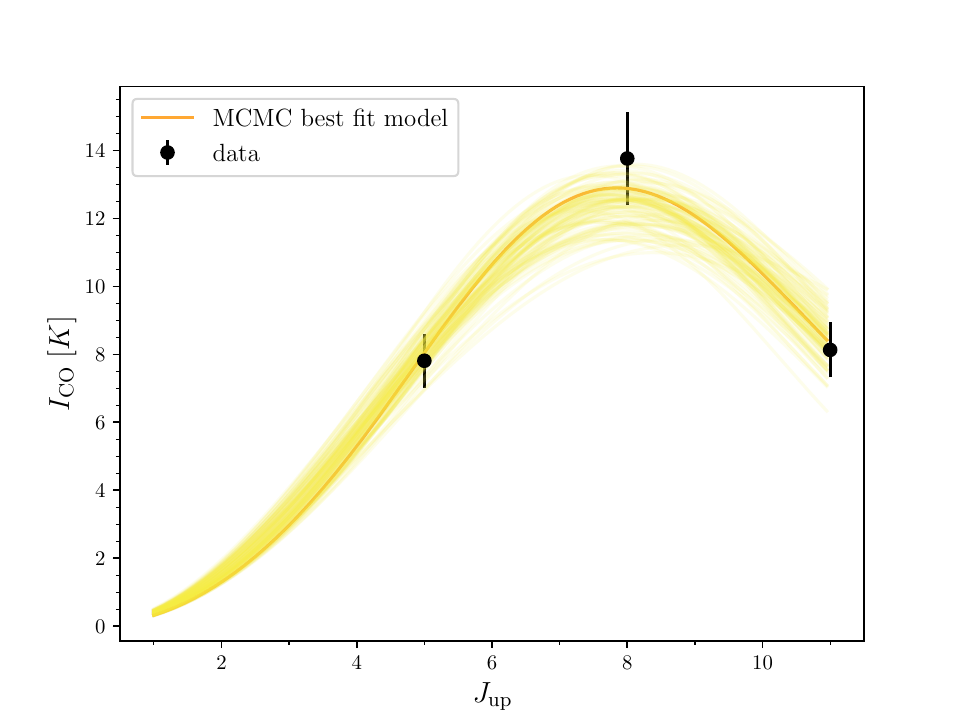}
\includegraphics[width=0.5\textwidth]{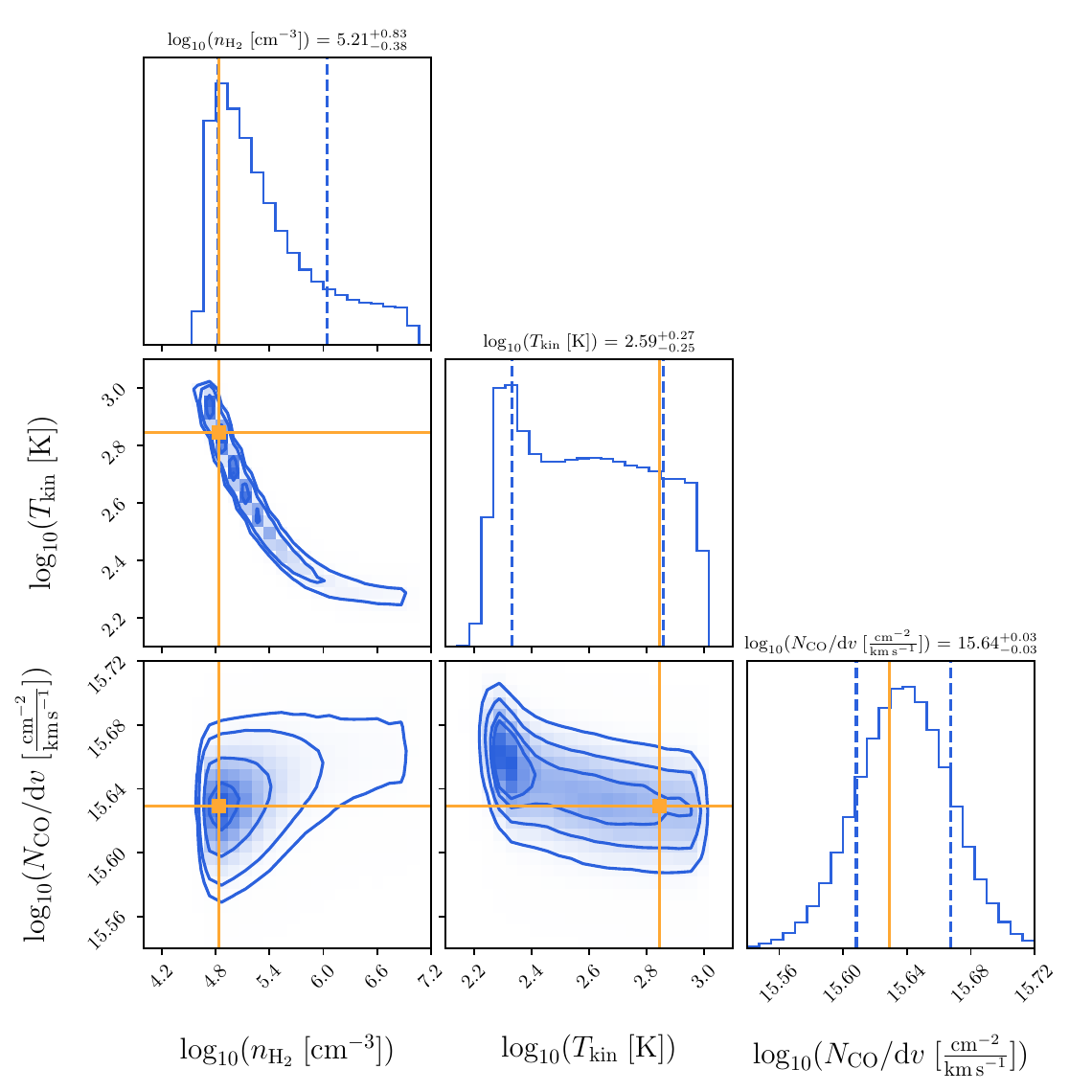}
\caption{RADEX modeling results with the MCMC sample of three free parameter spaces $n(\rm H_{2})$, $T_{\rm kin}$ and $N({\rm CO})$. \textbf{\textit{Top}}: Orange curve shows the best RADEX and emcee fitting to the intensity of three CO lines and yellow ones plot the 100 "good models" within the uncertainty range from the 16th to the 84th percentiles. \textbf{\textit{Bottom}}: Distribution of the best-fit parameters, in which the main diagonal shows the marginalized distributions of each parameter, with vertical dashed lines indicating the 16th and 84th percentiles, and the 2D plots show the correlation between each pair of parameters. The solid orange lines locate the best-fit model in the parameter space.}
\label{fig:NLTE_fit}
\end{figure}

The top panel of Figure \ref{fig:NLTE_fit} shows the best-fit model (orange line and yellow shaded area) to the observed data (black dots). The triangle plot in the bottom panel shows the distribution of the best-fit parameters. The vertical dashed lines mark the 16th and 84th percentile, which indicate the lower and upper values of the uncertainties, respectively. The posterior PDF for the gas density is peaked at $n_{\rm H_{2}}\simeq 1.5\times 10^{5}\,\rm cm^{-3}$, and $N(\rm CO)\simeq 10^{16}\,\rm cm^{-2}$, while the kinetic temperature is not constrained, with $T_{\rm kin}$ uncertainty ranges from $\approx 219 - 724\K$.  

Our constraint on the CO column density of $10^{16}-5\times 10^{17}\,\rm cm^{-2}$ and on the gas volume density of $10^{5}\,\rm cm^{-3}$ are similar to the ones of the neighbor Treasure Chest globule ($N(\rm CO)=8.5\times 10^{16}-4.9\times 10^{18}\,\rm cm^{-2}$ and $n_{\rm H_2} \sim 10^5 - 10^6 \cm^{-3}$, see \citealt{2019A&A...626A.131M}).

\section{Discussion}\label{sec:discussion}
\subsection{Motion of G287.76-0.87 Pillar}\label{sec:pillar_motion}
Determining the direction of motion relative to the illuminated source is important for determining the effects of stellar feedback on the pillar. To put the observed velocity of the [\textsc{Cii}] emission into context, we first assess its relation to the stars that are the source of ionization and wind energy.
The most luminous star in the region, $\eta$ Car itself, has a velocity of -19.7 \kms with respect to LSR \citep{smith04}. 
A survey of the radial velocities of 63 O-type stars in Carina has an average velocity of -11.0 \kms, with a wide dispersion of 9 \kms \citep{kiminki18}.
The pillar structure that is the subject of this paper is illuminated from the NW in the direction of Tr 16, which contains the most powerful Wolf-Rayet and O stars. The
average velocity of O stars in Tr 16 is -15.1 \kms \citep{kiminki18}.
It is important to note the wide dispersion in velocities among stars in each cluster and differences between clusters of order 10 \kms.

Now consider the molecular gas velocities. The MOPRA southern galactic plane survey shows a cold molecular gas traced by CO(1--0) emission at velocities around -20$\,$\kms \citep{2016MNRAS.456.2406R}.
The pillar G287.76-0.87 is a part of this complex of molecular gas
that is currently a work surface exposed to stellar radiation and winds. The dispersion of the velocity of CO is $\sim 6$ \kms, so a core formed from this cold gas could have a velocity anywhere from -17 to -23 \kms. 

\begin{figure}
    \centering
    \includegraphics[width=0.45\textwidth]{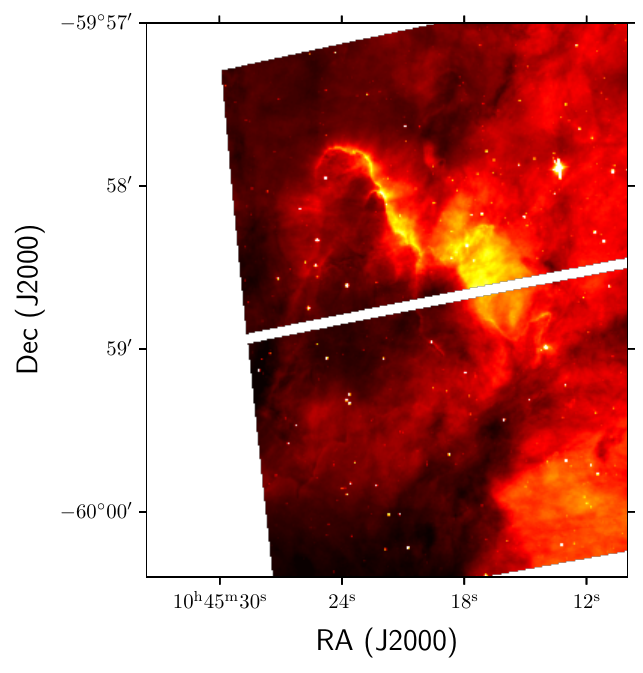}
    \caption{Hubble H$\alpha$ image of the pillar covering the same region as Figure \ref{fig:cut_PV}. The main body of the pillar is dark, apparently in silhouette. The rim and in particular the northwest edge (toward the luminous stars in Tr 16) is bright, indicating the pillar is closer to us than the illumination source.}
    \label{fig:HST_image}
\end{figure}

If the pillar gas started at the same -20$\,$\kms as the cold molecular gas nearby, and then the pressure from Tr 16 arises at a slightly greater distance than the pillar, it would be pushed toward more negative velocities, $v<-20\,$\kms. 
We measure a line-of-sight velocity of -14 \kms, which is contrary to the direction of expansion based on the assumptions of the
previous sentence. However, the present-day cold molecular gas in the vicinity may not be representative of the initial conditions for the pillar because the portion of the cloud that comprises the pillar may have been part of the parent clump that formed Tr 16. If that is the case, then we should compare its velocity with that of the stars, and the acceleration would lead to $v<-15.1$ \kms. 

Nevertheless, the observed radial velocity of the pillar still indicates that it is moving away from us by $\sim 1$ \kms (averaged over the pillar). In addition, this is only the radial component; the full 3-D velocity depends on the angle between the ray from the pillar toward Tr 16 and the line of sight. This angle could be quite large because the illumination appears in the sky's plane. Therefore, it is ambiguous to know whether the pillar is pushed away or toward us.

The {\it Hubble} image in Figure \ref{fig:HST_image} shows a sharp, bright, exposed rim to the NW, with a dark core, qualitatively suggesting that the illumination source is approximately the same distance as the core but perhaps somewhat further away so that a ray from Tr 16 to the molecular core lands on the NW side of the core just on the backside.

\subsection{Stellar feedback on G287.76-0.87 pillar}
As discussed in Section \ref{sec:pillar_motion}, the interpretation of the G287.76-0.87 motion is uncertain. As for a demonstration, consider that the pillar is accelerated by 1$\,$\kms, the kinetic energy required is $10^{45}\,$ erg due to a mass of order $10^2$ M$_\odot$. With a projected separation between the pillar and Tr 16 of 10 pc, and a pillar diameter of 0.7 pc, only 0.03\% of the stellar output is intercepted by it. Thus the energy output from the cluster would need to be of order $10^{48}$ erg over a spherical shell at 10 pc. The extant stars could produce such energies. Indeed, the pressure that an \textsc{Hii} region (i.e., hot gas) powered by an illumination source is $P_{\rm ext,HII}/k_{\rm B}=\sqrt{3Q/4\pi r^{3}\alpha_{r}}T_{\rm HII}$ (see Eq. 2 in \citealt{2009ApJ...693.1696H}) with $\alpha_{r}\simeq 4\times 10^{-13}$ the recombination coefficient, $Q=9\times 10^{50}\,\rm s^{-1}$ the ionizing photons per second. For a $T_{\rm HII}=10^{4}\,\rm K$, $P_{\rm ext,HII}/k_{\rm B}\sim 10^{6}\,\rm K\,cm^{-3}$. The equivalent energy of the cluster over a spherical shell at $10\,\rm pc$ is of the order $10^{49}\,\rm erg$. 

The photons from the radiation source Tr 16 also compress the pillar with a pressure of $P_{\rm ext,rad}=L_{\rm bol}/(4\pi r^{2} c)$, in the limit where all photons incident on the pillar are absorbed by it and the dust and gas are coupled \citep{tielens83}. Adopting the bolometric luminosity of $\simeq 10^{7}\,L_{\odot}$ (see e.g., \citealt{2006MNRAS.367..763S,2009ApJ...693.1696H}), the radiation pressure exerted on the pillar is about the same as the hot gas pressure as $P_{\rm ext,rad}/k_{\rm B}\sim 10^{6}\,\rm K\,cm^{-3}$.

As the cloud contracts, the internal pressure due to the thermal gas acts outward, causing the cloud to re-expand. Taking into account the hydrostatic pressure by nonthermal gas motion, the internal gas pressure of $P_{\rm int,th}/k_{\rm B}=nT_{\rm eff}$ with $T_{\rm eff}$ the effective gas temperature. As the velocity dispersion $\sigma = 0.92-1.18\,$\kms (Figure \ref{fig:spectra}), the representative effective temperature $T^{\rm average}_{\rm eff}=(0.92^{2}-1.18^{2})\times 10^{10} (\mu m_{\rm H}/k_{\rm B}$)$\,$K with $\mu=2.8$; thus, the average internal gas pressure is $P^{\rm average}_{\rm int,th}/k_{\rm B} \sim 10^{7}\,\rm K\,cm^{-3}$. As a consistency check, the thermal pressure of the gas could be estimated via the empirical relation to the radiation field as $P^{\rm PDR}_{\rm int,th}/k_{\rm B} = 2.1\times 10^{4} G_{\rm UV}^{0.9}$ resulted from the PDR models (\citealt{2018A&A...618A..53W}). For $G_{\rm UV} \sim 1000\times G_{0}$ at the location of our pillar (\citealt{2013A&A...554A...6R}), $P^{\rm PDR,average}_{\rm int,th}/k_{\rm B}\sim 10^{7}\, \K \cm^{-3}$. This internal pressure is about an order of magnitude higher than the external pressure of hot gas and radiation. This could be due to (1) the overestimation of $T_{\rm eff}$ (an average over integrated spectrum that could not represent the local variation) and (2) the gas heated externally and internally. Nevertheless, this simply implies that the pillar is supported against the feedback from these external pressures, which seems consistent with a small relative velocity with respect to $\eta$~Carina, and is plausible for the strong turbulent motion in the tail as seen in [\textsc{Oi}] as the gas is leaked through it (the gas density is lower at the tail).

Interestingly, the stellar wind pressure at 10$\,$pc is $P_{w}/k_{\rm B}\simeq 10^{9}\,\rm K\,cm^{-3}$ at 10$\,$pc assuming a standard mass-loss rate of $\sim 10^{-3}\,M_{\odot}\,\rm yr^{-1}$ (see \citealt{2009ApJ...693.1696H}), which is much higher than all above pressures. This implies that the stellar wind either escaped (leaked) through the holes (lowest density) to sculpt the farther distant ISM or was dissipated, or the wind luminosity is overestimated. In reality, a combination of these three factors might be the case. The former is phenomenally similar to the expansion of gas around the Orion Nebula Cluster \citep{pabst19} or 30 Doradus (\citealt{2021A&A...649A.175M}), and could be plausible for the giant bubble around Tr 16. The latter phenomenon could be due to the shocked gas in which the stellar wind loses its kinetic energy. A more systematic study of the velocities of the Carina pillars could yield a real determination of the feedback of the star cluster onto massive cores near one of the most powerful stellar clusters of the Milky Way.

\section{Summary}\label{sec:sum}

In this paper, we studied the kinematics of the pillar G287.76-0.87 in the southern region of the Carina Nebula, using the observational data from SOFIA/GREAT, including CO(5-4), CO(8-7), CO(11-10), and [\textsc{Oi}] 63$\,\mum$. Our main findings are summarized as follows
\begin{enumerate}
    \item We analyzed the channel maps and the resolved-spectral lines and showed that the LOS velocity of the pillar is -14$\km\,\s^{-1}$, which is more redshifted than the components of the Tr 16 cluster. The optical image reveals that the rim and the edge toward the luminous stars in Tr 16 are bright, indicating that the pillar is closer to us than the illumination source. Thus, the LOS velocity contradiction might be solely due to the project effect. 
    
    \item We made the moment maps of the four lines. The zeroth moment maps show that the pillar's northwest region (toward $\eta$~ Carlina) shows strong emission because of the radiative excitation and/or interaction with the \textsc{Hii} expansion shell. The [\textsc{Oi}] 63$\,\mum$ line partially correlates very well with the PAH continuum emission, and CO lines rise behind the atomic line, which illustrates the structure of the PDR. Second-moment maps show that the motion in the tail is more turbulent.
    
    \item We extracted the position-velocity (PV) diagram along several directions and indicated that the \textsc{Hii} expansion shell influences the gas motion in the northwest-to-southeast direction (Cut 1 in Figure \ref{fig:cut_PV}), while the gas is less dynamic in the perpendicular direction (cut 2 in Figure \ref{fig:cut_PV}). The diagram through the southernmost (cut 3 in Figure \ref{fig:cut_PV}) indicates a diffuse medium. 
    
    \item We constrained the physical properties of the G287.76-0.87 pillar under LTE and non-LTE conditions. For LTE, our method estimated the CO column density of $N_{\rm CO} \sim 2\times 10^{16} -5\times 10^{17}\cm^{-2}$. For non-LTE, we used RADEX with MCMC to estimate the gas number density, kinetic temperature, and CO column density. These estimations yield $n_{\rm H_{2}}\simeq 1.5\times 10^{5}\cm^{-3}$ and $N_{\rm CO}\simeq 10^{16}\cm^{-2}$.
    
    \item We discussed that the internal pressure of the gas within the G287.76-0.87 is sufficiently high that this pillar is hard to be pushed by the external hot gas and radiation pressures from Tr 16.
\end{enumerate}

{\it Acknowledgments}:
Based on observations made with the NASA/DLR Stratospheric Observatory for Infrared Astronomy (SOFIA). SOFIA is jointly operated by the Universities Space Research Association, Inc. (USRA), under NASA contract NNA17BF53C, and the Deutsches SOFIA Institut (DSI) under DLR contract 50 OK 0901 to the University of Stuttgart.

\software{GILDAS/CLASS (\citealt{2005sf2a.conf..721P}, GILDAS team 2013), APLpy (\citealt{2012ascl.soft08017R}), RADEX (\citealt{2007A&A...468..627V}), pvextractor (\citealt{2016ascl.soft08010G}), kalibrate (\citealt{2012A&A...542L...4G}), emcee (\citealt{2013PASP..125..306F}), pyradex \hyperlink{https://github.com/keflavich/pyradex}{(https://github.com/keflavich/pyradex})}

\newpage
\appendix
\section{Appendix A: Chanel maps of CO}
\begin{figure*}[!ht]
\centering
\includegraphics[width=0.9\textwidth]{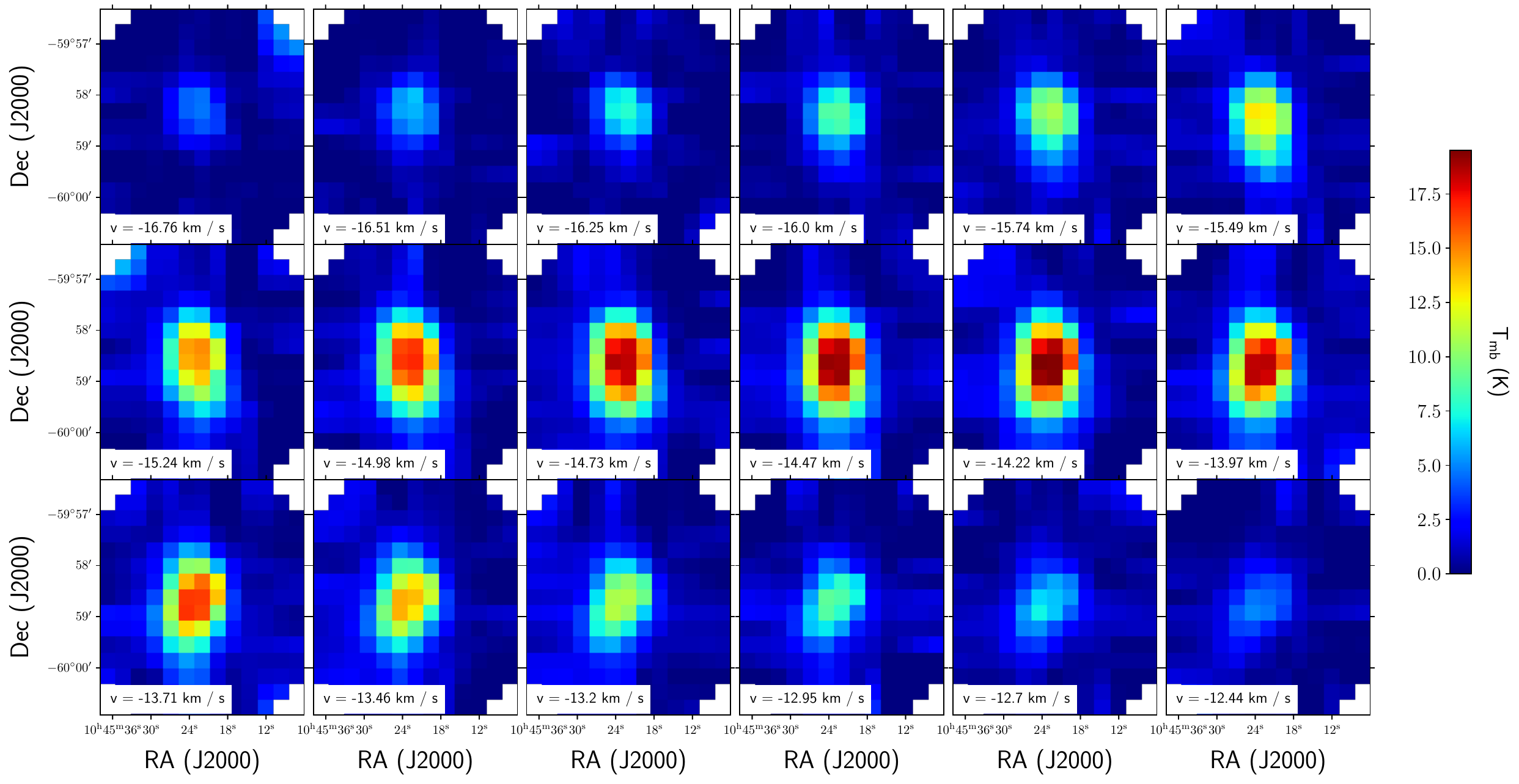}
\caption{Channel maps of CO(5-4).}
\label{fig:channelmap_app2}
\end{figure*}

\begin{figure*}[!ht]
\centering
\includegraphics[width=0.9\textwidth]{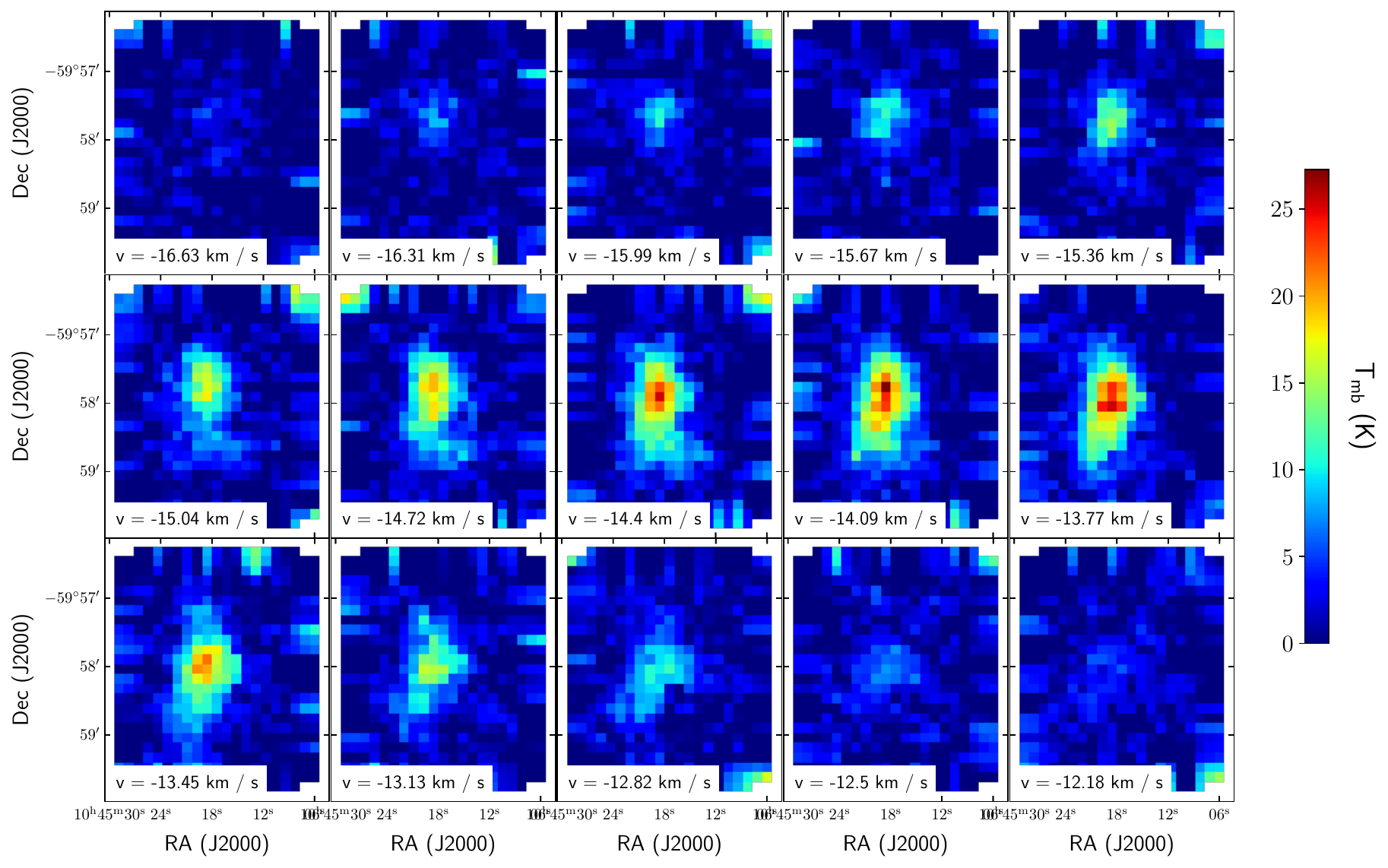}
\caption{Channel maps of CO(8-7).}
\label{fig:channelmap_app3}
\end{figure*}

\begin{figure*}
\centering
\includegraphics[width=0.9\textwidth]{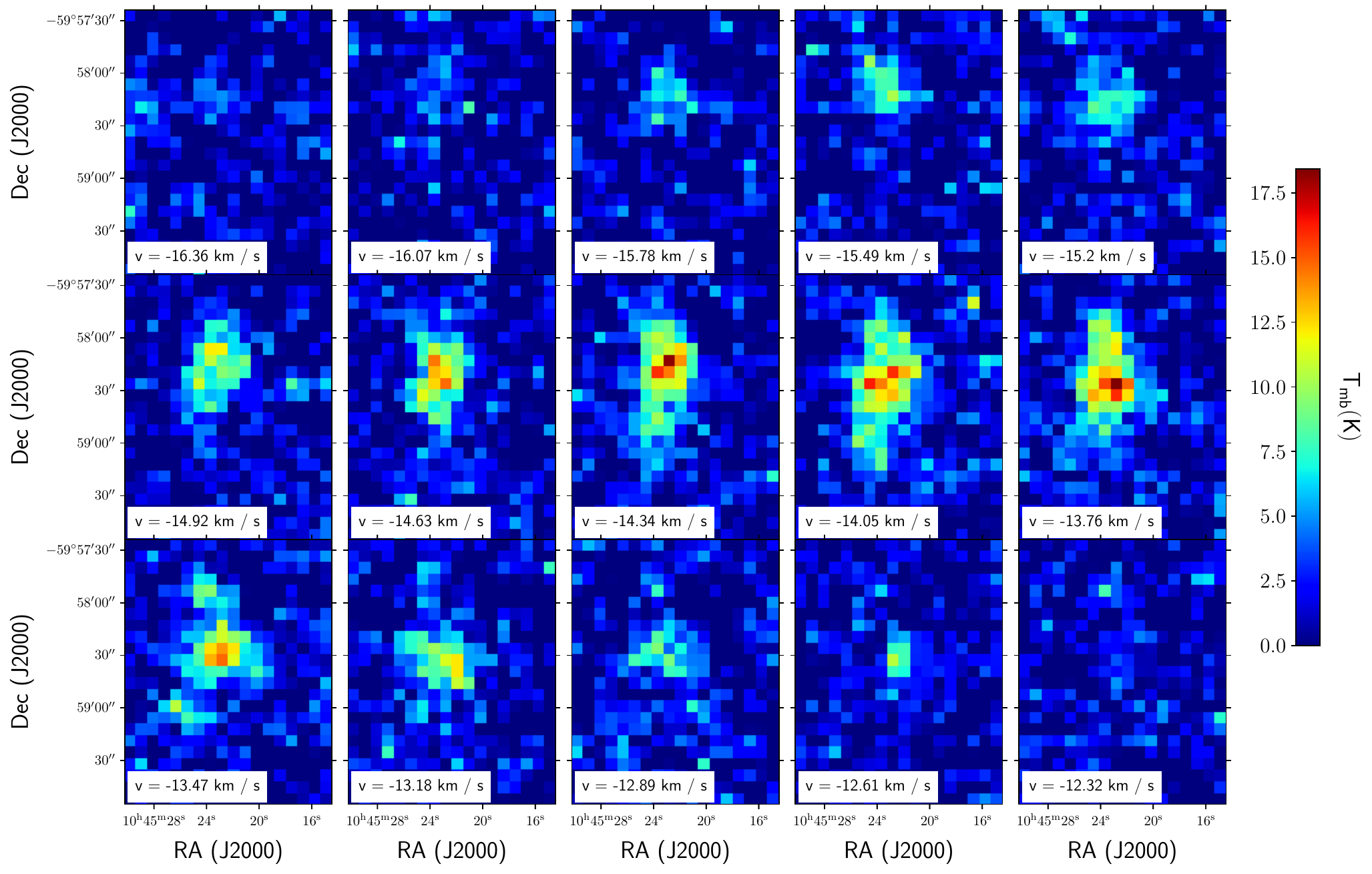}
\caption{Channel maps of CO(10-11).}
\label{fig:channelmap_app4}
\end{figure*}

Figures \ref{fig:channelmap_app2}-\ref{fig:channelmap_app4} show the channel maps of the $^{12}$CO(5-4), $^{12}$CO(8-7) and $^{12}$CO(11-10). Owing to the larger beam sizes, $^{12}$CO maps are able to spatially resolve the Pillar structure. However, the bulk velocity is also $\simeq -14\,\rm km\,s^{-1}$ as seen in [\textsc{OI}] (see Figure \ref{fig:channelmap}).

\newpage

\section{Appendix B: Moment maps of CO transition lines}
\begin{figure*}
\centering
\includegraphics[width=0.43\textwidth]{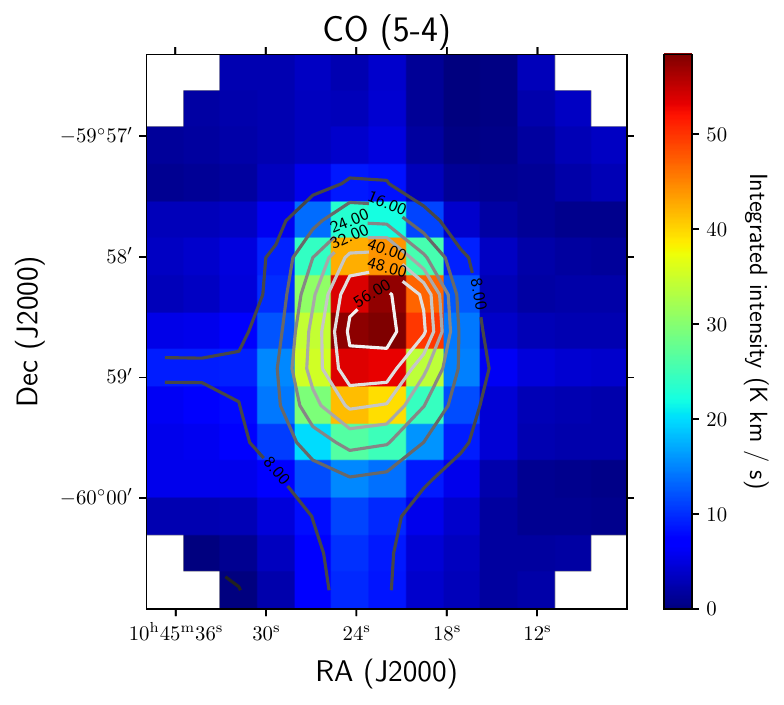}
\includegraphics[width=0.45\textwidth]{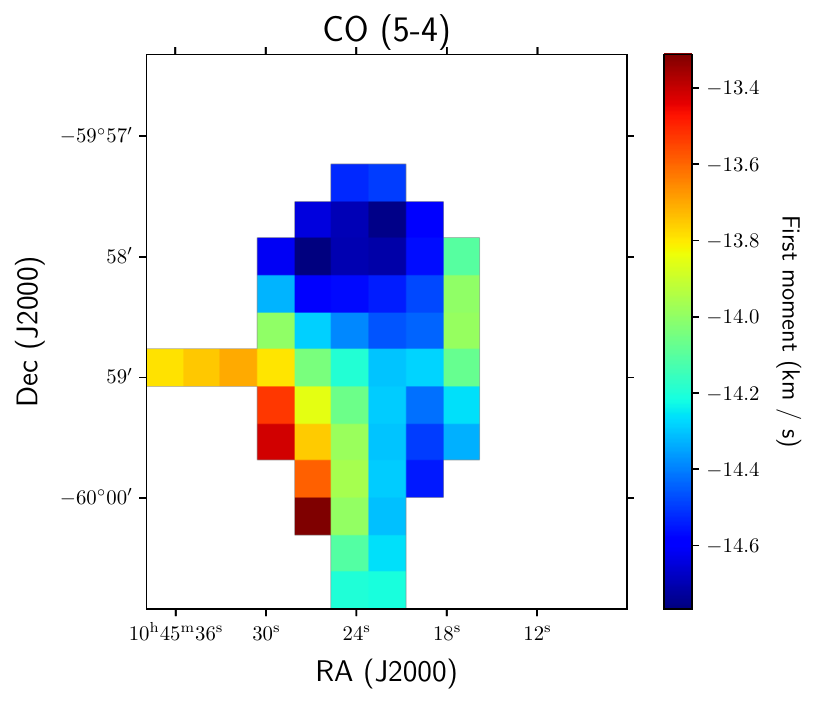}
\includegraphics[width=0.45\textwidth]{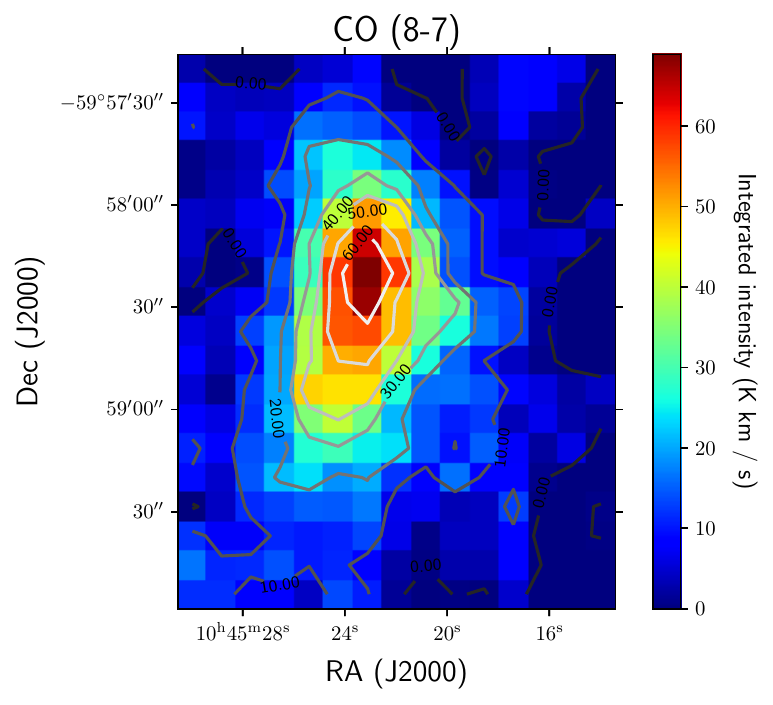}
\includegraphics[width=0.45\textwidth]{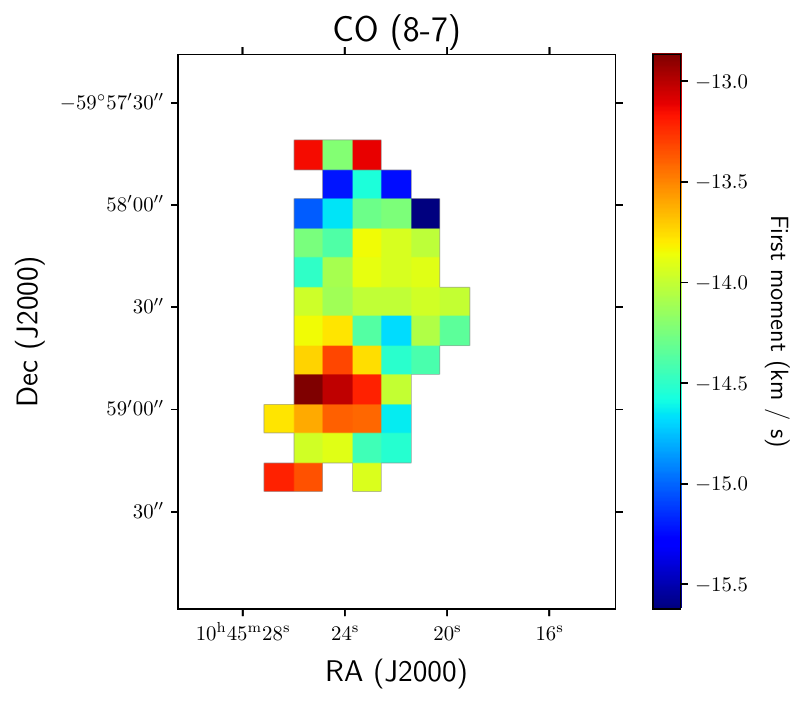}
\includegraphics[width=0.45\textwidth]{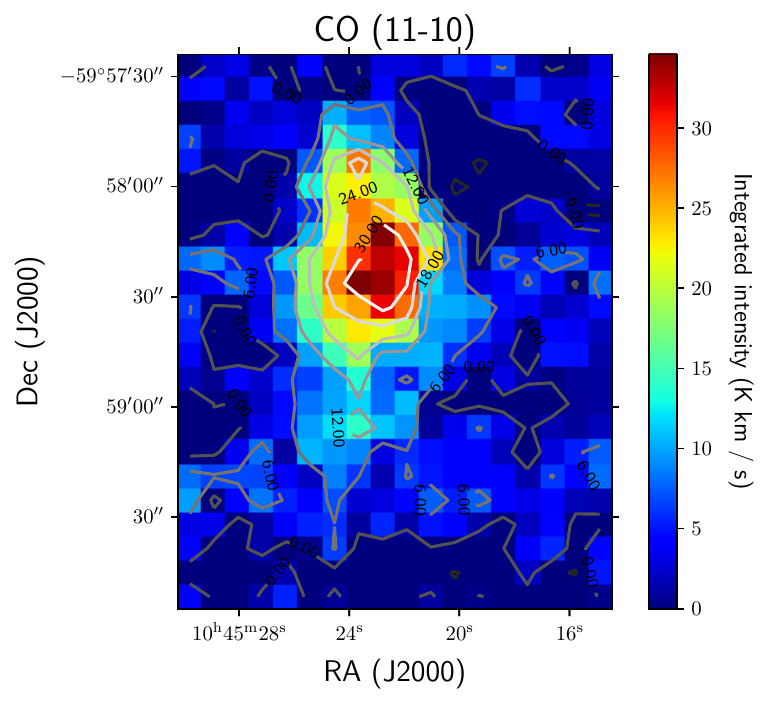}
\includegraphics[width=0.47\textwidth]{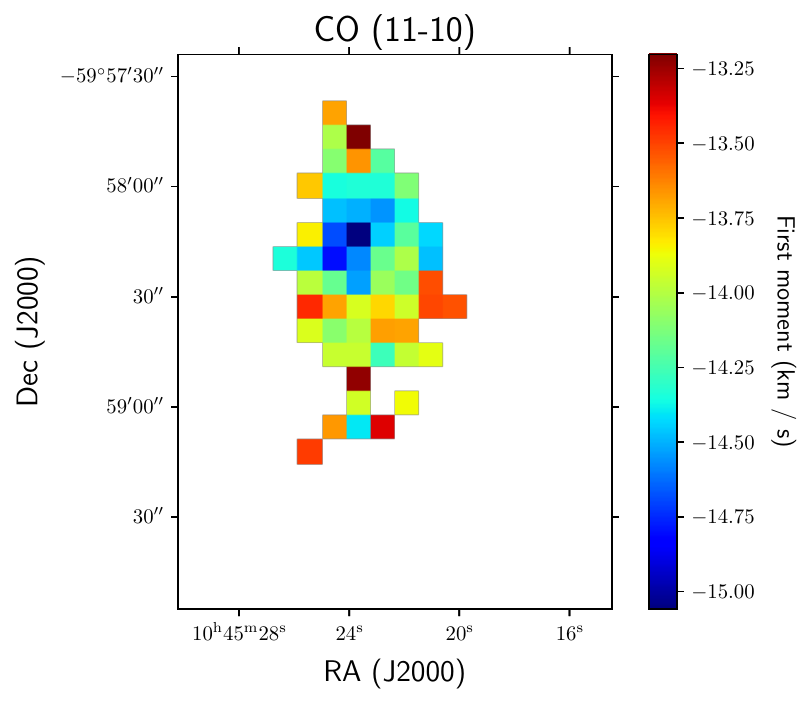}
\caption{Integrated intensity maps (left panel) and mean velocity maps (right panel) of CO(5-4), CO(8-7), and CO(11-10) lines. The contours show the level of the integrated intensity.}
\label{fig:moment01_app}
\end{figure*}

The left panels in Figure \ref{fig:moment01_app} show the moment-0 maps of $^{12}$CO(5-4), $^{12}$CO(8-7) and $^{12}$CO(11-10), respectively from top to bottom. Similarly, the right panels show the moment-1 maps. These $^{12}$CO lines are bright in emission, but their maps are unable to resolve the Pillar structure.

\begin{thebibliography}{55}
\expandafter\ifx\csname natexlab\endcsname\relax\def\natexlab#1{#1}\fi

\bibitem[{Akimkin {et~al.}(2013)Akimkin, Zhukovska, Wiebe, Semenov,
  Pavlyuchenkov, Vasyunin, Birnstiel, \& Henning}]{2013ApJ...766....8A}
Akimkin, V., Zhukovska, S., Wiebe, D., {et~al.} 2013, \apj, 766, 8

\bibitem[{Allamandola {et~al.}(1993)Allamandola, Sandford, Tielens, \&
  Herbst}]{1993Sci...260...64A}
Allamandola, L.~J., Sandford, S.~A., Tielens, A. G. G.~M., \& Herbst, T.~M.
  1993, Science (ISSN 0036-8075), 260, 64

\bibitem[{Allamandola {et~al.}(1985)Allamandola, Tielens, \&
  Barker}]{1985ApJ...290L..25A}
Allamandola, L.~J., Tielens, A. G. G.~M., \& Barker, J.~R. 1985, \apj, 290, L25

\bibitem[{Brunner {et~al.}(2018)Brunner, Maercker, Mecina, Khouri, \&
  Kerschbaum}]{Brunner:2018fn}
Brunner, M., Maercker, M., Mecina, M., Khouri, T., \& Kerschbaum, F. 2018,
  Astronomy and Astrophysics, 614, A17

\bibitem[{Casassus {et~al.}(2007)Casassus, Nyman, Dickinson, \&
  Pearson}]{2007MNRAS.382.1607C}
Casassus, S., Nyman, L.-{\AA}., Dickinson, C., \& Pearson, T.~J. 2007, Monthly
  Notices of the Royal Astronomical Society, 382, 1607

\bibitem[{Castellanos {et~al.}(2011)Castellanos, Casassus, Dickinson, Vidal,
  Paladini, Cleary, Davies, Davis, White, \& Taylor}]{2011MNRAS.411.1137C}
Castellanos, P., Casassus, S., Dickinson, C., {et~al.} 2011, \mnras, 411, 1137

\bibitem[{Cherchneff(2011)}]{2011EAS....46..177C}
Cherchneff, I. 2011, EAS Publications Series, 46, 177

\bibitem[{Cherchneff(2012)}]{2012A&A...545A..12C}
Cherchneff, I. 2012, Astronomy and Astrophysics, 545, A12

\bibitem[{Chiu {et~al.}(2006)Chiu, Hoang, {Dinh-V-Trung}, Lim, Kwok, Hirano, \&
  Muthu}]{2006ApJ...645..605C}
Chiu, P.-J., Hoang, C.-T., {Dinh-V-Trung}, {et~al.} 2006, \apj, 645, 605

\bibitem[{Decin {et~al.}(2011)Decin, Royer, Cox, Vandenbussche, Ottensamer,
  Blommaert, Groenewegen, Barlow, Lim, Kerschbaum, Posch, \&
  Waelkens}]{Decin:2011dh}
Decin, L., Royer, P., Cox, N. L.~J., {et~al.} 2011, Astronomy and Astrophysics,
  534, A1

\bibitem[{Dehaes {et~al.}(2007)Dehaes, Groenewegen, Decin, Hony, Raskin, \&
  Blommaert}]{2007MNRAS.377..931D}
Dehaes, S., Groenewegen, M. A.~T., Decin, L., {et~al.} 2007, Monthly Notices of
  the Royal Astronomical Society, 377, 931

\bibitem[{Dickinson {et~al.}(2018)Dickinson, Ali-Ha{\"\i}moud, Barr,
  Battistelli, Bell, Bernstein, Casassus, Cleary, Draine, G{\'e}nova-Santos,
  Harper, Hensley, Hill-Valler, Hoang, Israel, Jew, Lazarian, Leahy, Leech,
  L{\'o}pez-Caraballo, McDonald, Murphy, Onaka, Paladini, Peel, Perrott,
  Poidevin, Readhead, Rubi{\~n}o-Mart{\'\i}n, Taylor, Tibbs, Todorovi{\'c}, \&
  Vidal}]{Dickinson:2018ix}
Dickinson, C., Ali-Ha{\"\i}moud, Y., Barr, A., {et~al.} 2018, New Astronomy
  Reviews, 80, 1

\bibitem[{Draine(2006)}]{2006ApJ...636.1114D}
Draine, B.~T. 2006, \apj, 636, 1114

\bibitem[{Draine \& Hensley(2012)}]{2012ApJ...757..103D}
Draine, B.~T., \& Hensley, B. 2012, \apj, 757, 103

\bibitem[{Draine \& Lazarian(1998)}]{1998ApJ...508..157D}
Draine, B.~T., \& Lazarian, A. 1998, \apj, 508, 157

\bibitem[{Draine \& Li(2007)}]{2007ApJ...657..810D}
Draine, B.~T., \& Li, A. 2007, \apj, 657, 810

\bibitem[{Draine {et~al.}(2007)Draine, Dale, Bendo, Gordon, Smith, Armus,
  Engelbracht, Helou, Kennicutt, Li, Roussel, Walter, Calzetti, Moustakas,
  Murphy, Rieke, Bot, Hollenbach, Sheth, \& Teplitz}]{2007ApJ...663..866D}
Draine, B.~T., Dale, D.~A., Bendo, G., {et~al.} 2007, \apj, 663, 866

\bibitem[{Duley \& Grishko(2001)}]{2001ApJ...554L.209D}
Duley, W.~W., \& Grishko, V.~I. 2001, The Astrophysical Journal, 554, L209

\bibitem[{Dwek(2016)}]{2016ApJ...825..136D}
Dwek, E. 2016, \apj, 825, 136

\bibitem[{Fleming \& Stone(2003)}]{2003ApJ...585..908F}
Fleming, T., \& Stone, J.~M. 2003, \apj, 585, 908

\bibitem[{Geballe {et~al.}(1989)Geballe, Noll, Whittet, \&
  Waters}]{1989ApJ...340L..29G}
Geballe, T.~R., Noll, K.~S., Whittet, D. C.~B., \& Waters, L. B. F.~M. 1989,
  Astrophysical Journal, 340, L29

\bibitem[{Goldreich \& Scoville(1976)}]{1976ApJ...205..144G}
Goldreich, P., \& Scoville, N. 1976, \apj, 205, 144

\bibitem[{Gu{\'e}lin {et~al.}(2018)Gu{\'e}lin, Patel, Bremer, Cernicharo,
  Castro-Carrizo, Pety, Fonfr{\'\i}a, Ag{\'u}ndez, Santander-Garc{\'\i}a,
  Quintana-Lacaci, Velilla~Prieto, Blundell, \& Thaddeus}]{2018A&A...610A...4G}
Gu{\'e}lin, M., Patel, N.~A., Bremer, M., {et~al.} 2018, Astronomy and
  Astrophysics, 610, A4

\bibitem[{Hensley \& Draine(2017)}]{2017ApJ...836..179H}
Hensley, B.~S., \& Draine, B.~T. 2017, \apj, 836, 179

\bibitem[{Hensley {et~al.}(2016)Hensley, Draine, \&
  Meisner}]{2016ApJ...827...45H}
Hensley, B.~S., Draine, B.~T., \& Meisner, A.~M. 2016, \apj, 827, 45

\bibitem[{Hoang(2017)}]{Hoang:2017kg}
Hoang, T. 2017, The Astrophysical Journal, 847, 0

\bibitem[{Hoang {et~al.}(2010)Hoang, Draine, \& Lazarian}]{Hoang:2010jy}
Hoang, T., Draine, B.~T., \& Lazarian, A. 2010, \apj, 715, 1462

\bibitem[{Hoang {et~al.}(2018)Hoang, Lan, Vinh, \& Kim}]{Hoang:2018td}
Hoang, T., Lan, N.-Q., Vinh, N.-A., \& Kim, Y.-J. 2018, arXiv:1803.11028

\bibitem[{Hoang {et~al.}(2011)Hoang, Lazarian, \& Draine}]{2011ApJ...741...87H}
Hoang, T., Lazarian, A., \& Draine, B.~T. 2011, \apj, 741, 87

\bibitem[{Hoang {et~al.}(2016)Hoang, Vinh, \& Quynh~Lan}]{2016ApJ...824...18H}
Hoang, T., Vinh, N.~A., \& Quynh~Lan, N. 2016, \apj, 824, 18

\bibitem[{Knapp {et~al.}(1995)Knapp, Bowers, Young, \&
  Phillips}]{1995ApJ...455..293K}
Knapp, G.~R., Bowers, P.~F., Young, K., \& Phillips, T.~G. 1995, Astrophysical
  Journal v.455, 455, 293

\bibitem[{Knapp {et~al.}(1999)Knapp, Young, \& Crosas}]{1999A&A...346..175K}
Knapp, G.~R., Young, K., \& Crosas, M. 1999, Astronomy and Astrophysics, 346,
  175

\bibitem[{Kogut {et~al.}(1996)Kogut, Banday, Bennett, Gorski, Hinshaw, Smoot,
  \& Wright}]{Kogut:1996p5293}
Kogut, A., Banday, A.~J., Bennett, C.~L., {et~al.} 1996, \apjl, 464, L5

\bibitem[{Kwok(1975)}]{1975ApJ...198..583K}
Kwok, S. 1975, Astrophysical Journal, 198, 583

\bibitem[{Leger \& Puget(1984)}]{1984A&A...137L...5L}
Leger, A., \& Puget, J.-L. 1984, A\&A, 137, L5

\bibitem[{Leitch {et~al.}(1997)Leitch, Readhead, Pearson, \&
  Myers}]{Leitch:1997p7359}
Leitch, E.~M., Readhead, A. C.~S., Pearson, T.~J., \& Myers, S.~T. 1997, \apjl,
  486, L23

\bibitem[{Li(2004)}]{2004ASPC..309..417L}
Li, A. 2004, Astrophysics of Dust, 309, 417

\bibitem[{Li \& Draine(2001)}]{Li:2001p4761}
Li, A., \& Draine, B.~T. 2001, \apj, 550, L213

\bibitem[{Li {et~al.}(2013)Li, Liu, \& Jiang}]{2013ApJ...777..111L}
Li, A., Liu, J.~M., \& Jiang, B.~W. 2013, The Astrophysical Journal, 777, 111

\bibitem[{McDonald {et~al.}(2011)McDonald, Zijlstra, Sloan, \&
  Matsuura}]{2011ASPC..445..241M}
McDonald, I., Zijlstra, A.~A., Sloan, G.~C., \& Matsuura, M. 2011, in Why
  Galaxies Care about AGB Stars II: Shining Examples and Common Inhabitants.
  Proceedings of a conference held at University Campus, 241--

\bibitem[{Meeus {et~al.}(2001)Meeus, Waters, Bouwman, van~den Ancker, Waelkens,
  \& Malfait}]{2001A&A...365..476M}
Meeus, G., Waters, L. B. F.~M., Bouwman, J., {et~al.} 2001, A\&A, 365, 476

\bibitem[{Menten {et~al.}(2006)Menten, Reid, Kr{\"u}gel, Claussen, \&
  Sahai}]{2006A&A...453..301M}
Menten, K.~M., Reid, M.~J., Kr{\"u}gel, E., Claussen, M.~J., \& Sahai, R. 2006,
  Astronomy and Astrophysics, 453, 301

\bibitem[{Newville {et~al.}(2014)Newville, Stensitzki, Allen, \&
  Ingargiola}]{newville_2014_11813}
Newville, M., Stensitzki, T., Allen, D.~B., \& Ingargiola, A. 2014, {LMFIT:
  Non-Linear Least-Square Minimization and Curve-Fitting for Python¶}

\bibitem[{{Planck Collaboration} {et~al.}(2016){Planck Collaboration}, Adam,
  Ade, Aghanim, \& et~al.}]{2016A&A...594A..10P}
{Planck Collaboration}, Adam, R., Ade, P. A.~R., Aghanim, N., \& et~al. 2016,
  Astronomy and Astrophysics, 594, A10

\bibitem[{{Planck Collaboration} {et~al.}(2011){Planck Collaboration}, Ade,
  Alves, \& et~al.}]{PlanckCollaboration:2011hw}
{Planck Collaboration}, Ade, P. A.~R., Alves, M. I.~R., \& et~al. 2011, A\&A,
  536, A20

\bibitem[{Ramos-Larios {et~al.}(2011)Ramos-Larios, Phillips, \&
  Cuesta}]{2011MNRAS.411.1245R}
Ramos-Larios, G., Phillips, J.~P., \& Cuesta, L.~C. 2011, Monthly Notices of
  the Royal Astronomical Society, 411, 1245

\bibitem[{Roberge {et~al.}(1995)Roberge, Hanany, \&
  Messinger}]{1995ApJ...453..238R}
Roberge, W.~G., Hanany, S., \& Messinger, D.~W. 1995, \apj, 453, 238

\bibitem[{Sahai {et~al.}(1989)Sahai, Claussen, \& Masson}]{1989A&A...220...92S}
Sahai, R., Claussen, M.~J., \& Masson, C.~R. 1989, Astronomy and Astrophysics
  (ISSN 0004-6361), 220, 92

\bibitem[{Sahai {et~al.}(2011)Sahai, Claussen, Schnee, Morris, \&
  S{\'a}nchez~Contreras}]{2011ApJ...739L...3S}
Sahai, R., Claussen, M.~J., Schnee, S., Morris, M.~R., \&
  S{\'a}nchez~Contreras, C. 2011, The Astrophysical Journal Letters, 739, L3

\bibitem[{Sandell {et~al.}(2011)Sandell, Weintraub, \&
  Hamidouche}]{2011ApJ...727...26S}
Sandell, G., Weintraub, D.~A., \& Hamidouche, M. 2011, The Astrophysical
  Journal, 727, 26

\bibitem[{Seok \& Li(2017)}]{2017ApJ...835..291S}
Seok, J.~Y., \& Li, A. 2017, \apj, 835, 291

\bibitem[{Smith {et~al.}(2007)Smith, Draine, Dale, \&
  et~al.}]{2007ApJ...656..770S}
Smith, J.-D.~T., Draine, B.~T., Dale, D.~A., \& et~al. 2007, \apj, 656, 770

\bibitem[{Testi {et~al.}(2015)Testi, Perez, Jimenez-Serra, Hoare, Boley,
  Bourke, Brucato, Caselli, Chandler, Isella, Lazio, Palumbo, Podio, Remijan,
  Tarter, \& Wilner}]{2015aska.confE.117T}
Testi, L., Perez, L., Jimenez-Serra, I., {et~al.} 2015, in Proceedings of
  Advancing Astrophysics with the Square Kilometre Array (AASKA14). 9 -13 June,
  117

\bibitem[{Tielens(2008)}]{2008ARA&A..46..289T}
Tielens, A. G. G.~M. 2008, \araa, 46, 289

\bibitem[{Tielens {et~al.}(1987)Tielens, Tielens, Seab, Seab, Hollenbach,
  Hollenbach, McKee, \& McKee}]{1987ApJ...319L.109T}
Tielens, A. G. G.~M., Tielens, A. G. G.~M., Seab, C.~G., {et~al.} 1987,
  Astrophysical Journal, 319, L109

\bibitem[{200(2003)}]{2003agbs.conf.....H}
 2003, {Asymptotic giant branch stars}

\bibitem[{{Ag{\'u}ndez} \& {Cernicharo}(2006)}]{2006ApJ...650..374A}
{Ag{\'u}ndez}, M., \& {Cernicharo}, J. 2006, \apj, 650, 374

\bibitem[{{Burke} \& {Silk}(1974)}]{1974ApJ...190....1B}
{Burke}, J.~R., \& {Silk}, J. 1974, \apj, 190, 1

\bibitem[{{Crosas} \& {Menten}(1997)}]{1997ApJ...483..913C}
{Crosas}, M., \& {Menten}, K.~M. 1997, \apj, 483, 913

\bibitem[{{De Beck} {et~al.}(2010){De Beck}, {Decin}, {de Koter}, {Justtanont},
  {Verhoelst}, {Kemper}, \& {Menten}}]{2010A&A...523A..18D}
{De Beck}, E., {Decin}, L., {de Koter}, A., {et~al.} 2010, \aap, 523, A18

\bibitem[{{Decin} {et~al.}(2010){Decin}, {De Beck}, {Br{\"u}nken},
  {M{\"u}ller}, {Menten}, {Kim}, {Willacy}, {de Koter}, \&
  {Wyrowski}}]{2010A&A...516A..69D}
{Decin}, L., {De Beck}, E., {Br{\"u}nken}, S., {et~al.} 2010, \aap, 516, A69

\bibitem[{{Draine} \& {Salpeter}(1979)}]{1979ApJ...231...77D}
{Draine}, B.~T., \& {Salpeter}, E.~E. 1979, \apj, 231, 77

\bibitem[{{Gilman}(1972)}]{1972ApJ...178..423G}
{Gilman}, R.~C. 1972, \apj, 178, 423

\bibitem[{{Hoang} \& {Tram}(2019)}]{2019ApJ...877...36H}
{Hoang}, T., \& {Tram}, L.~N. 2019, \apj, 877, 36

\bibitem[{{Hoang} {et~al.}(2019){Hoang}, {Tram}, {Lee}, \&
  {Ahn}}]{2019NatAs.tmp..319H}
{Hoang}, T., {Tram}, L.~N., {Lee}, H., \& {Ahn}, S.-H. 2019, Nature Astronomy,
  319

\bibitem[{{Jaeger} {et~al.}(1998){Jaeger}, {Molster}, {Dorschner}, {Henning},
  {Mutschke}, \& {Waters}}]{Jaeger_1998}
{Jaeger}, C., {Molster}, F.~J., {Dorschner}, J., {et~al.} 1998, \aap, 339, 904

\bibitem[{{Jones} \& {Spitzer}(1967)}]{1967ApJ...147..943J}
{Jones}, R.~V., \& {Spitzer}, Lyman, J. 1967, \apj, 147, 943

\bibitem[{{Justtanont} {et~al.}(2012){Justtanont}, {Khouri}, {Maercker},
  {Alcolea}, {Decin}, {Olofsson}, {Sch{\"o}ier}, {Bujarrabal}, {Marston},
  {Teyssier}, {Cernicharo}, {Dominik}, {de Koter}, {Melnick}, {Menten},
  {Neufeld}, {Planesas}, {Schmidt}, {Szczerba}, \&
  {Waters}}]{2012A&A...537A.144J}
{Justtanont}, K., {Khouri}, T., {Maercker}, M., {et~al.} 2012, \aap, 537, A144

\bibitem[{{Knapp} {et~al.}(1998){Knapp}, {Young}, {Lee}, \&
  {Jorissen}}]{1998ApJS..117..209K}
{Knapp}, G.~R., {Young}, K., {Lee}, E., \& {Jorissen}, A. 1998, \apjs, 117, 209

\bibitem[{{Krueger} {et~al.}(1994){Krueger}, {Gauger}, \&
  {Sedlmayr}}]{1994A&A...290..573K}
{Krueger}, D., {Gauger}, A., \& {Sedlmayr}, E. 1994, \aap, 290, 573

\bibitem[{{Kwan} \& {Linke}(1982)}]{1982ApJ...254..587K}
{Kwan}, J., \& {Linke}, R.~A. 1982, \apj, 254, 587

\bibitem[{{Kwok}(2004)}]{Kwok_2004}
{Kwok}, S. 2004, \nat, 430, 985

\bibitem[{{Li} {et~al.}(2016){Li}, {Millar}, {Heays}, {Walsh}, {van Dishoeck},
  \& {Cherchneff}}]{2016A&A...588A...4L}
{Li}, X., {Millar}, T.~J., {Heays}, A.~N., {et~al.} 2016, \aap, 588, A4

\bibitem[{{Maercker} {et~al.}(2016){Maercker}, {Danilovich}, {Olofsson}, {De
  Beck}, {Justtanont}, {Lombaert}, \& {Royer}}]{2016A&A...591A..44M}
{Maercker}, M., {Danilovich}, T., {Olofsson}, H., {et~al.} 2016, \aap, 591, A44

\bibitem[{{Mamon} \& {Soneira}(1982)}]{1982ApJ...255..181M}
{Mamon}, G.~A., \& {Soneira}, R.~M. 1982, \apj, 255, 181

\bibitem[{{Mathis} \& {Whiffen}(1989)}]{1989ApJ...341..808M}
{Mathis}, J.~S., \& {Whiffen}, G. 1989, \apj, 341, 808

\bibitem[{{Matthews} {et~al.}(2015){Matthews}, {G{\'e}rard}, \& {Le
  Bertre}}]{2015MNRAS.449..220M}
{Matthews}, L.~D., {G{\'e}rard}, E., \& {Le Bertre}, T. 2015, \mnras, 449, 220

\bibitem[{{Menten} {et~al.}(2012){Menten}, {Reid}, {Kami{\'n}ski}, \&
  {Claussen}}]{2012A&A...543A..73M}
{Menten}, K.~M., {Reid}, M.~J., {Kami{\'n}ski}, T., \& {Claussen}, M.~J. 2012,
  \aap, 543, A73

\bibitem[{{Millar} {et~al.}(2000){Millar}, {Herbst}, \&
  {Bettens}}]{2000MNRAS.316..195M}
{Millar}, T.~J., {Herbst}, E., \& {Bettens}, R.~P.~A. 2000, \mnras, 316, 195

\bibitem[{{Sahai} \& {Chronopoulos}(2010)}]{2010ApJ...711L..53S}
{Sahai}, R., \& {Chronopoulos}, C.~K. 2010, \apj, 711, L53

\bibitem[{{Tielens}(1983)}]{1983ApJ...271..702T}
{Tielens}, A.~G.~G.~M. 1983, \apj, 271, 702

\bibitem[{{Tram} {et~al.}(2018){Tram}, {Lesaffre}, {Cabrit}, \&
  {Nhung}}]{2018arXiv180801439T}
{Tram}, L.~N., {Lesaffre}, P., {Cabrit}, S., \& {Nhung}, P.~T. 2018, arXiv
  e-prints, arXiv:1808.01439

\bibitem[{{Winters} {et~al.}(1994){Winters}, {Dominik}, \&
  {Sedlmayr}}]{Winters_1994}
{Winters}, J.~M., {Dominik}, C., \& {Sedlmayr}, E. 1994, \aap, 288, 255

\end{thebibliography}


\newpage
\begin{thebibliography}
\expandafter\ifx\csname natexlab\endcsname\relax\def\natexlab#1{#1}\fi
\bibitem[Anderson et al.(2019)]{2019ApJ...882...11A} Anderson, L.~D., Makai, Z., Luisi, M., et al.\ 2019, \apj, 882, 11. doi:10.3847/1538-4357/ab1c59

\bibitem[de Jong et al.(1980)]{1980A&A....91...68D} de Jong, T., Boland, W., \& Dalgarno, A.\ 1980, \aap, 91, 68

\bibitem[Draine(2011)]{2011piim.book.....D} Draine, B.~T.\ 2011, Physics of the Interstellar and Intergalactic Medium by Bruce T. Draine. Princeton University Press, 2011. ISBN: 978-0-691-12214-4

\bibitem[Dur{\'a}n et al.(2021)]{duran21} Dur{\'a}n, C.~A., G{\"u}sten, R., Risacher, C., et al.\ 2021, IEEE Transactions on Terahertz Science and Technology, 11, 194. doi:10.1109/TTHZ.2020.3042714

\bibitem[{{Elmegreen} \& {Lada}(1977)}]{1977ApJ...214..725E}
{Elmegreen}, B.~G., \& {Lada}, C.~J. 1977, \apj, 214, 725,
  \doi{10.1086/155302}
  
\bibitem[Foreman-Mackey et al.(2013)]{2013PASP..125..306F} Foreman-Mackey, D., Hogg, D.~W., Lang, D., et al. 2013, \pasp, 125, 306. doi:10.1086/670067

\bibitem[Ginsburg et al.(2016)]{2016ascl.soft08010G} Ginsburg, A., Robitaille, T., \& Beaumont, C.\ 2016, Astrophysics Source Code Library. ascl:1608.010

\bibitem[Goldsmith(2019)]{2019ApJ...887...54G} Goldsmith, P.~F.\ 2019, \apj, 887, 54. doi:10.3847/1538-4357/ab535e

\bibitem[Goodman \& Weare(2010)]{2010CAMCS...5...65G} Goodman, J. \& Weare, J.\ 2010, Communications in Applied Mathematics and Computational Science, 5, 65. doi:10.2140/camcos.2010.5.65

\bibitem[GILDAS (2013)]{GILDAS2013} GILDAS Team (2013), GILDAS: Grenoble Image and Line Data Analysis Software, Astrophysics Source Code Library, ascl:1305.010.


\bibitem[Guan et al.(2012)]{2012A&A...542L...4G} Guan, X., Stutzki, J., Graf, U.~U., et al.\ 2012, \aap, 542, L4. 

\bibitem[{{Harper-Clark} \& {Murray}(2009)}]{2009ApJ...693.1696H}
{Harper-Clark}, E., \& {Murray}, N. 2009, \apj, 693, 1696,
  \doi{10.1088/0004-637X/693/2/1696}
  
\bibitem[Hester et al.(1996)]{1996AJ....111.2349H} Hester, J.~J., Scowen, P.~A., Sankrit, R., et al. 1996, \aj, 111, 2349. doi:10.1086/117968

\bibitem[Hollenbach \& Tielens(1997)]{1997ARA&A..35..179H} Hollenbach, D.~J. \& Tielens, A.~G.~G.~M.\ 1997, \araa, 35, 179. doi:10.1146/annurev.astro.35.1.179

\bibitem[Hsieh et al.(2012)]{2012ApJS..203...23H} Hsieh, B.-C., Wang, W.-H., Hsieh, C.-C., et al.\ 2012, \apjs, 203, 23. doi:10.1088/0067-0049/203/2/23

\bibitem[Kahn(1954)]{1954BAN....12..187K} Kahn, F.~D.\ 1954, \bain, 12, 187

\bibitem[Kavak et al.(2022)]{kavak22} Kavak, {\" U}., Bally, J., Goicoechea, J. R.,  Pabst, C. H. M.,  van der Tak, F. F. S.,  Tielens, A. G. G. M. 2022, \aap, 663, A117. doi:10.1051/0004-6361/202243332 


\bibitem[Kiminki and Smith(2018)]{kiminki18} {{Kiminki}, M. M. and {Smith}, N.} 2018, \mnras, 477, 2068

\bibitem[Kirsanova et al.(2020)]{2020MNRAS.497.2651K} Kirsanova, M.~S., Ossenkopf-Okada, V., Anderson, L.~D., et al.\ 2020, \mnras, 497, 2651. doi:10.1093/mnras/staa2142


\bibitem[Kirsanova et al.(2023)]{kirsanova23} Kirsanova, M. S., Pavlyuchenkov, Y. N. ,  Olofsson, A. O. H.,  Semenov, D. A.,  Punanova, A. F. 2023, \mnras, 520, 751. doi:10.1093/mnras/stac3737


\bibitem[McMullin et al.(2007)]{2007ASPC..376..127M} McMullin, J.~P., Waters, B., Schiebel, D., et al.\ 2007, Astronomical Data Analysis Software and Systems XVI, 376, 127

\bibitem[{{Melnick} {et~al.}(2021){Melnick}, {Tenorio-Tagle}, \&
  {Telles}}]{2021A&A...649A.175M}
{Melnick}, J., {Tenorio-Tagle}, G., \& {Telles}, E. 2021, \aap, 649, A175,
  \doi{10.1051/0004-6361/201937268}
  
\bibitem[Mihalas(1978)]{1978stat.book.....M} Mihalas, D.\ 1978, San Francisco: W.H. Freeman, 1978

\bibitem[Mookerjea et al.(2019)]{2019A&A...626A.131M} Mookerjea, B., Sandell, G., G{\"u}sten, R., et al.\ 2019, \aap, 626, A131. doi:10.1051/0004-6361/201935482

\bibitem[Pabst et al.(2019)]{pabst19} Pabst, C. et al. 2019, Nature, 565, 618

\bibitem[P\'erez-Beaupuits et al.(2012)]{perez12} P\'erez-Beaupuits, J. P. et al. 2012, \aap, 542, L13

\bibitem[{{Pety}(2005)}]{2005sf2a.conf..721P}
{Pety}, J. 2005, in SF2A-2005: Semaine de l'Astrophysique Francaise, ed.
  F.~{Casoli}, T.~{Contini}, J.~M. {Hameury}, \& L.~{Pagani}, 721

\bibitem[{{Rathborne} {et~al.}(2004){Rathborne}, {Brooks}, {Burton}, {Cohen},
  \& {Bontemps}}]{2004A&A...418..563R}
{Rathborne}, J.~M., {Brooks}, K.~J., {Burton}, M.~G., {Cohen}, M., \&
  {Bontemps}, S. 2004, \aap, 418, 563, \doi{10.1051/0004-6361:20031631}

\bibitem[Rebolledo et al.(2016)]{2016MNRAS.456.2406R} Rebolledo, D., Burton, M., Green, A., et al.\ 2016, \mnras, 456, 2406. doi:10.1093/mnras/stv2776

  
\bibitem[Rebolledo et al.(2020)]{rebolledo20} Rebolledo, D. et al. 2020, ApJ, 891, 113

\bibitem[Risacher et al.(2018)]{2018JAI.....740014R} Risacher, C., G{\"u}sten, R., Stutzki, J., et al.\ 2018, Journal of Astronomical Instrumentation, 7, 1840014. doi:10.1142/S2251171718400147

\bibitem[Robitaille \& Bressert(2012)]{2012ascl.soft08017R} Robitaille, T. \& Bressert, E.\ 2012, Astrophysics Source Code Library. ascl:1208.017

\bibitem[Robitaille et al.(2016)]{2016ascl.soft09017R} Robitaille, T., Ginsburg, A., Beaumont, C., et al.\ 2016, Astrophysics Source Code Library. ascl:1609.017

\bibitem[{{Roccatagliata} {et~al.}(2013){Roccatagliata}, {Preibisch}, {Ratzka},
  \& {Gaczkowski}}]{2013A&A...554A...6R}
{Roccatagliata}, V., {Preibisch}, T., {Ratzka}, T., \& {Gaczkowski}, B. 2013,
  \aap, 554, A6, \doi{10.1051/0004-6361/201321081}
  
\bibitem[Schneider et al.(2012)]{2012A&A...542L..18S} Schneider, N., G{\"u}sten, R., Tremblin, P., et al.\ 2012, \aap, 542, L18. doi:10.1051/0004-6361/201218917

\bibitem[{{Smith}(2006)}]{2006MNRAS.367..763S}
{Smith}, N. 2006, \mnras, 367, 763, \doi{10.1111/j.1365-2966.2006.10007.x}

\bibitem[Smith (2004)]{smith04} Smith, N. 2004, MNRAS, 351, L15

\bibitem[Spitzer(1998)]{1998ppim.book.....S} Spitzer, L.\ 1998, Physical Processes in the Interstellar Medium, by Lyman Spitzer, pp. 335. ISBN 0-471-29335-0. Wiley-VCH , May 1998., 335

\bibitem[Smith(2005)]{2005AJ....129..888S} Smith, N.,Stassun, Keivan G., Bally, John, \ 2005, \aj, 129, 888. 
doi: 10.1086/427249 

\bibitem[Str{\"o}mgren(1939)]{1939ApJ....89..526S} Str{\"o}mgren, B.\ 1939, \apj, 89, 526. doi:10.1086/144074

\bibitem[Thompson et al.(2004)]{2004A&A...414.1017T} Thompson, M.~A., White, G.~J., Morgan, L.~K., et al.\ 2004, \aap, 414, 1017. doi:10.1051/0004-6361:20031680

\bibitem[Tielens(1983)]{tielens83} Tielens, A.~G.~G.~M.\ 1983, \aap, 119, 177

\bibitem[van der Tak et al.(2007)]{2007A&A...468..627V} van der Tak, F.~F.~S., Black, J.~H., Sch{\"o}ier, F.~L., et al.\ 2007, \aap, 468, 627

\bibitem[White(1993)]{1993A&A...274L..33W} White, G.~J.\ 1993, \aap, 274, L33

\bibitem[White \& Gee(1986)]{1986A&A...156..301W} White, G.~J. \& Gee, G.\ 1986, \aap, 156, 301

\bibitem[{{Wu} {et~al.}(2018){Wu}, {Bron}, {Onaka}, {Le Petit}, {Galliano},
  {Languignon}, {Nakamura}, \& {Okada}}]{2018A&A...618A..53W}
{Wu}, R., {Bron}, E., {Onaka}, T., {et~al.} 2018, \aap, 618, A53, \doi{10.1051/0004-6361/201832595}

\bibitem[Yang et al.(2017)]{2017A&A...608A.144Y} Yang, C., Omont, A., Beelen, A., et al.\ 2017, \aap, 608, A144. \doi{10.1051/0004-6361/201731391}

\bibitem[Young et al.(2012)]{young12} Young, E.
T., and 40 coauthors, 2012, \apjl, 749, 17
\doi{10.1088/2041-8205/749/2/L17}

\end{thebibliography}
\end{document}